\documentclass[12pt,preprint]{aastex}

\shorttitle{Mid--IR emission of Seyfert galaxies.}
\shortauthors{Ramos Almeida et al.}

\begin{document}

\title{The mid--infrared emission of Seyfert galaxies.
 A new analysis of ISOCAM data\footnote{Based on observations with ISO, an 
ESA project with instruments funded by ESA Member States 
(especially the PI countries: France, Germany, the Netherlands and 
the United Kingdom) and with the participation of ISAS and NASA.}}

\author{C. Ramos Almeida\altaffilmark{1}, A.M. P\'{e}rez Garc\'\i a\altaffilmark{1},
J.A. Acosta-Pulido\altaffilmark{1}, and J.M. Rodr\'\i guez Espinosa\altaffilmark{1}}

\altaffiltext{1}{Instituto de Astrof\'\i sica de Canarias (IAC), 
              C/V\'\i a L\'{a}ctea, s/n, E-38200, La Laguna, Tenerife, Spain.
cra@iac.es, apg@iac.es, jap@iac.es, and jre@iac.es}

\begin{abstract}

We present mid--infrared data of a sample of 57 AGNs obtained with the instrument ISOCAM on board
the satellite ISO. The images were obtained through the LW2 (6.75~\micron) and LW7 (9.62~\micron)
filters. 
This is a new analysis of \citet{Clavel00} galaxy sample, which is divided into 26 type 1 ($\le$ 1.5) 
and 28 type 2 ($>$ 1.5) Seyfert galaxies, plus three QSOs. 
The spatial resolution of the images allow us to separate the nuclear and the extended
contributions to the total emission after decomposing the brightness profiles into different
morphological components. The most common components are a central point source (identified
as the active nucleus) and an exponential disk. In some cases a bulge, a bar or a ring are needed. 
The relative contribution of the nucleus to the total emission appears larger in Seyfert 1 
than in Seyfert 2. This result confirms that both types of Seyfert galaxies are 
different in the mid-infrared wavelength range and supports 
the existence of an structure which produces anisotropic emission in this wavelength range.
We have also explored correlations between the mid-infrared and the radio and X--ray wavelength ranges. 
The well established radio/infrared correlation is mantained in our sample for the global
emission of the galaxies. If only the nuclear infrared emission is considered
then a non--linear correlation is apparent in the luminosity--luminosity scatter diagram. 
The ratio between the intrinsic hard X--ray and the 
nuclear mid-infrared emission presents large scatter and slightly larger values for type 2 
Seyfert galaxies. These results seem to be consistent with the presence of a clumpy 
dusty torus surrounding the active nucleus.

\end{abstract}


\section{Introduction}
 
	Despite the broad variety of type of objects included under the denomination of Active Galactic
Nuclei (AGN), it seems to be possible to explain all of them under a common scenario, the so called 
Unification Models. The most successful models for the unification of Seyfert galaxies predict the 
existence of a blocking structure surrounding the nucleus \citep{Antonucci93}. 
This structure is likely a dusty torus, whose size
and structure is still a matter of debate \citep{Fritz06,Honig06}. 
According to this, type 1 and  2 Seyfert galaxies (hereafter Sy1 and Sy2,
respectively) are proposed as the same kind of objects but viewed at different angles: 
Sy1 are observed close to
face-on such that we have a direct view of the nucleus and the Broad emission Line Region (BLR), 
whereas the Sy2 are seen at an inclination such that our view is blocked by the 
optically thick dusty torus. 
The dust grains in the torus will absorb the UV photons from the central engine 
and after reprocessing the radiation will appear as strong emission in the infrared range. 
In particular, the mid-infrared emission is produced by a mixture of stochastic heating 
from Polycyclic Aromatic Hydrocarbon (PAHs) and thermal emission from very small dust grains at 
high temperatures. 
Active galaxies are significantly stronger radiators in the mid-infrared than non 
AGN-dominated galaxies \citep{Spinoglio89,Rowan89, Fadda98, Perez01}.
Hence, this spectral range seems a natural window in which to study the properties
of such a structure. 

	The mid-infrared spectra of Sy1 and Sy2 galaxies present important differences: whereas Sy1
spectra are characterized by a strong continuum with only weak emission features from PAH bands, most 
Sy2 display a weak continuum but very strong PAH emission bands \citep{Clavel00}. 
Despite the strong dilution by the nuclear continuum in the case of Sy1, both types 1 and 2 share 
similar PAH luminosities. 
This molecular emission results unrelated to the
nuclear activity, and arises in the interstellar medium of the underlying galactic bulge.
The majority of dusty torus models predict different mid--infrared spectra for type 1 and 2
objects. In case of Sy1 the silicate feature at 10~\micron\ is expected in emission, as recently
confirmed with instruments on board of the Spitzer satellite \citep{Siebenmorgen05}. In contrast,
for Sy2 this feature is observed in absorption \citep{Clavel00,Jaffe04}.

	In this paper, we analyze mid-infrared images of a sample of Seyfert galaxies. 
The data sample practically coincides with the previosly studied by
\citet{Clavel00}, in which the images are complemented by spectra
obtained with the instrument ISOPHOT--S \citep{Lemke96}.
Our main goal is to isolate the nuclear emission of the galaxies in the sample, 
and provide a better estimate of the active component.  
This new analysis of the data can be used for testing the unification models. 
In this line we have investigated correlations with photometric data obtained in different 
spectral ranges, namely  radio and hard X--rays.

\section{ISOCAM Data}

We analyzed the mid-infrared morphology of a sample of AGNs, mostly Seyfert galaxies. 
These data were taken as part of the ISO Guaranteed Program {\it Seyfert 1 and Seyfert 2} lead by 
J. Clavel.  
We have selected this dataset because it was taken in a homogeneous way, with all 
the exposures obtained in the same mode and keeping the same instrumental configuration
(pixel scale and filters).
Morphological classification, Seyfert type, spectroscopic redshifts
and observational data (exposure and observation time) are reported in Table \ref{info}.  
The sample was originally drawn from the CfA hard X-ray flux limited complete sample 
\citep{Piccinotti82}, but lacks the most well known objects (e.g. NGC~4151 or NGC~1068) 
which were obtained within the frame of other ISO guaranteed time programs. 
The sample is about equally divided into 
26 Sy1 ($\le$ 1.5) and 28 Sy2 ($>$ 1.5), plus a starburst galaxy and three QSOs. 
Taking into account only the Seyfert galaxies, the mean and {\it rms} of the redshift
distributions are 0.023 $\pm$ 0.014 and 0.016 $\pm$ 0.011 for Sy1 and Sy2, respectively. 
The mean values are not significatively different, considering  the width
of the distributions, clarifying that the results presented in this paper 
are not a consequence of differences between Sy1 and Sy2 redshift distributions.
In addition to this, we have compiled the total H magnitudes for all Seyferts in the sample from 
The Two Micron All Sky Survey (2MASS), in order to derive their luminosity distributions.
The mean log(L$_{TOT}$[H]) are 44.81 $\pm$ 0.39 and 44.72 $\pm$ 0.42, indicating that  
Sy1 and Sy2 luminosity, and consequently, mass distributions are very similar.
The images were obtained using ISOCAM, the mid-infrared camera \citep{Cesarsky96} 
on board of the ISO spacecraft \citep{Kessler96}.  
The ISOCAM images are formed by an array of 32x32 pixels, with a pixel size of 3 arcsec. 
The employed filters were LW2 (5-8.5 \micron, 
covering PAHs emission) and LW7 (8.5-10.7 \micron, including the silicate band), 
which provide an effective resolution on the difraction limit 
of FWHM=$3\farcs8$ and $4\farcs5$, respectively.
Reduced data were retrieved from the ISO data archive\footnote{http://www.iso.vilspa.esa.es/ida/}.

\section{Data Analysis}

\subsection{Brightness profile decomposition}

We performed an isophotal analysis of the images in both filters to obtain the surface 
brightness profiles, by azimuthally averaging over elliptical annuli. We fitted the isocontours 
of surface brightness with the ELLIPSE task in IRAF\footnote{IRAF is
distributed by the National Optical Astronomy Observatories, which
are operated by the Association of Universities for the Research in
Astronomy, Inc., under cooperative agreement with the National science
Foundation. http://iraf.noao.edu/}, which employes the algorithm described in \citet{Jedrzejewski87}.
The spatial interval between two consecutive isophotes chosen was equal to the pixel size in all cases.
Given the limited spatial resolution and the relatively small field of view of the camera, 
the brightness profiles are limited to a maximum of 15 data points,
which means a relatively poor sampling of the profile.  
We subtracted the sky background emission, estimated as the median background value for each image. 
As a first approach, we assume that the profiles are the sum of two contributions: 
a nuclear component modeled with a Gaussian PSF of FWHM equals to 2~pixels\footnote{Here we assumed 
that the nominal spatial resolution cannot be achieved due to insufficient sampling of the 
PSF. The width of the  profile for a point source should correspond to the minimum sampling criterium, 
{\it i.e.} 2 pixels or equivalently 6\arcsec.},  
and central amplitude as free parameter plus an exponential component with 
two free parameters (central amplitude and length scale). Using this simple model we 
were able to reproduce correctly the brightness profiles
for about half of the galaxies in our sample. In the rest of cases we could not 
succesfully reproduce the brightness profiles,and extra components had to be introduced 
in order to improve the fit. The extra components are a bulge described  as a Sersic's 
law; a ring described by a Gaussian profile \citep{Buta96}:	
$$ I_{ring}(r) = I_{ring}^0 exp[-{1\over2}({r-r_{ring} \over l_{ring}})^{2}] ,$$ 	
where I$_{ring}^0$ is the amplitude, r$_{ring}$ is the center, and l$_{ring}$ is the width; or a flat 
bar profile \citep{Prieto97}:	
$$ I_{bar}(r) = {I_{bar}^0 \over 1+exp[(r-r_{bar})/l_{bar}]} , $$ 	
where I$_{bar}^0$ is the amplitude, r$_{bar}$ is the lenght, and l$_{bar}$ is the downward gradient. 

In all cases, the fits were performed with the criterium of minimum number of components and 
consistency between both filters.
%We also took special care with the unresolved component, in order
%to characterize as well as possible the nuclear emission of each galaxy.
Whenever bars or rings were employed in the fitting, we also looked at the ellipticity 
and orientation of the isophotes searching for abrupt changes coincident with the 
position of these components. However, the limited resolution
of our imaging prevents a fully reliable component identification.  
In addition to this, we used as a guideline the morphology of the galaxies in the visible (WFPC2/HST) 
from \citet{Malkan98}
and in the near-infrared (NICMOS/HST) from \citet{Hunt04}, although in
some cases we fit nuclear bars that do not agree with the optical classification, or rings that 
reproduce spiral arms, as in the case of NGC~1241. The maximum number of fitted parameters in our fits
was seven. 
The fitting procedure looks for minimization of the Chi-Square function formed by the squared 
difference between the observed and model profile, both in logarithmic scale. 
The multidimensional minimization of the merit function is reached using the downhill simplex method.   
This is the same fitting technique employed in \citet{Melo02} in the study of NGC~253. 
In Table \ref{params} we report the fit parameters and their corresponding errors for 54 objects of the sample, 
in both LW2 and LW7 filters. Errors were estimated using a bootstrapping
technique. This technique is based on a Monte Carlo simulation: new brightness profiles are obtained
by perturbation of the measured profile, using a normal distribution of the same width as the 
error of the measured point. New fit parameters are determined for each one of the simulated profiles
(30 simulations) and the uncertainty of each parameter is computed as the standard deviation 
of the resulting values.     
We calculated the fluxes integrating
all the emission contained in each fitted component and in the total fit, for both LW2 and LW7 filters.
As an additional check point, we verified that the
resultant flux from the total fit was at least $\sim 95$\% of the total measured flux. 
All fluxes and its errors are reported in Table \ref{fluxes}. 
We found that bulges are only needed in the case of three galaxies: Mrk~3, MGC6-30-15,
and Mrk~841, which are morphologically classified as S0 or 
elliptical (see Table \ref{info}). 
27 galaxies were fitted using a PSF plus an exponential component, 14 with PSF plus exponential component
plus bar, 4 with exponential component plus bar, 3 with PSF plus exponential component plus ring, 2 
with PSF plus bar plus ring, and one with exponential component
only. The galaxies NGC~7592 and ESO137-G34
can not be fitted because of their double nucleus, NGC~5929 because of the proximity of its companion 
(the starburst galaxy NGC~5930), and the QSO HS1700+6416 because of their high $z$=2.736.
For these four galaxies without profile decomposition, total fluxes were
obtained by means of aperture photometry, using the PHOT task, within the IRAF environment.  
 
\begin{figure*}
\centering
\includegraphics[width=18cm,height=20cm]{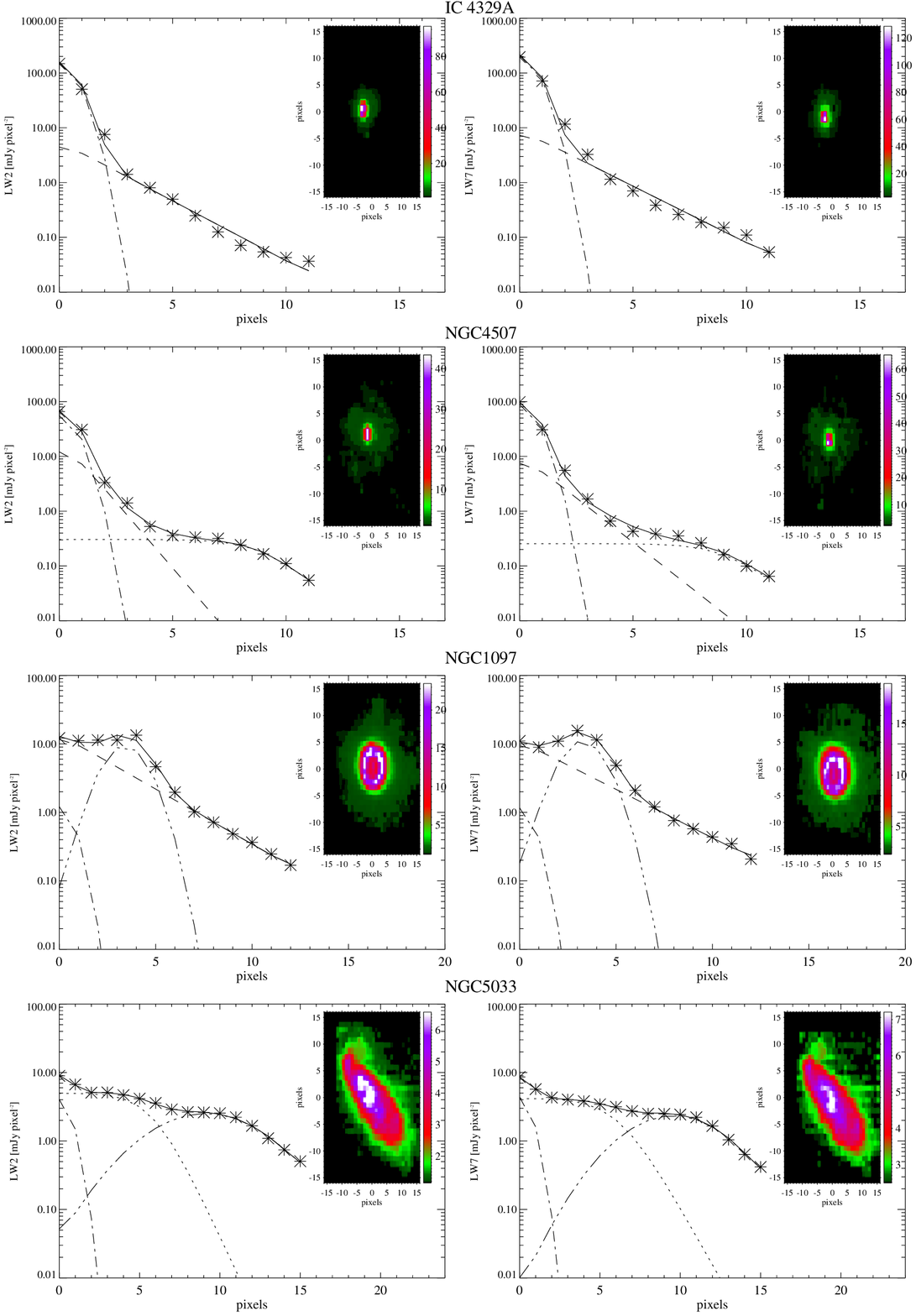}
\figcaption{\scriptsize{Brightness profiles of the Sy1 galaxy IC~4329 in both
LW2 and LW7 filters, fitted by an exponential
component (dashed line) plus a Gaussian PSF (dot-dashed line). Filled line is the final fit.
The same for the Sy2 galaxy NGC~4507, fitted by an 
exponential component (dashed line) plus a Gaussian PSF (dot-dashed line) and a flat bar (dotted line); 
for the Sy1 galaxy NGC~1097, fitted by an 
exponential component (dashed line) plus a Gaussian PSF (dot-dashed line) and a nuclear ring 
(double-dot-dashed line); and finally for the Sy2 galaxy NGC~5033, fitted by a 
Gaussian PSF (dot-dashed line) plus a flat bar (dotted line) and a ring (double-dot-dashed line).
Figures 1.1 - 1.50 are available in the electronic edition of the journal.} 
\label{ic4329a}}
\end{figure*}

We show in Fig.~\ref{ic4329a} examples of 
brightness profile decomposition, fitting different morphological components, in both LW2 (left panels) and LW7
(right panels) filters. The origin of X-axis corresponds to the center of fitted isophotes, 
and distance from this origin is given in pixels, being the pixel size of 3\arcsec. 

\clearpage

%\clearpage

%\begin{figure*}
%\centering
%\includegraphics[width=9cm,angle=90]{f2_color.eps}
%\figcaption{\scriptsize{Same as in Fig.~\ref{ic4329a}, but for the Sy2 galaxy NGC~4507, fitted by an 
%exponential component (dashed line) plus a Gaussian PSF (dot-dashed line) and a flat bar (dotted line). } 
%\label{ngc4507}}
%\end{figure*}

%\begin{figure*}
%\centering
%\includegraphics[width=9cm,angle=90]{f3_color.eps}
%\figcaption{\scriptsize{Same as in Fig.~\ref{ic4329a}, but for the Sy1 galaxy NGC~1097, fitted by an 
%exponential component (dashed line) plus a Gaussian PSF (dot-dashed line) and a nuclear ring 
%(double-dot-dashed line).} 
%\label{ngc1097}}
%\end{figure*}

%\begin{figure*}
%\centering
%\includegraphics[width=9cm,angle=90]{f4_color.eps}
%\figcaption{\scriptsize{Same as in Fig.~\ref{ic4329a}, but for the Sy2 galaxy NGC~5033, fitted by a 
%Gaussian PSF (dot-dashed line) plus a flat bar (dotted line) and a ring (double-dot-dashed line).} 
%\label{ngc5033}}
%\end{figure*}

\subsection{Comparison with previous studies}

\begin{figure}[!h]
\centering
{\par
\includegraphics[width=5.5cm,angle=90]{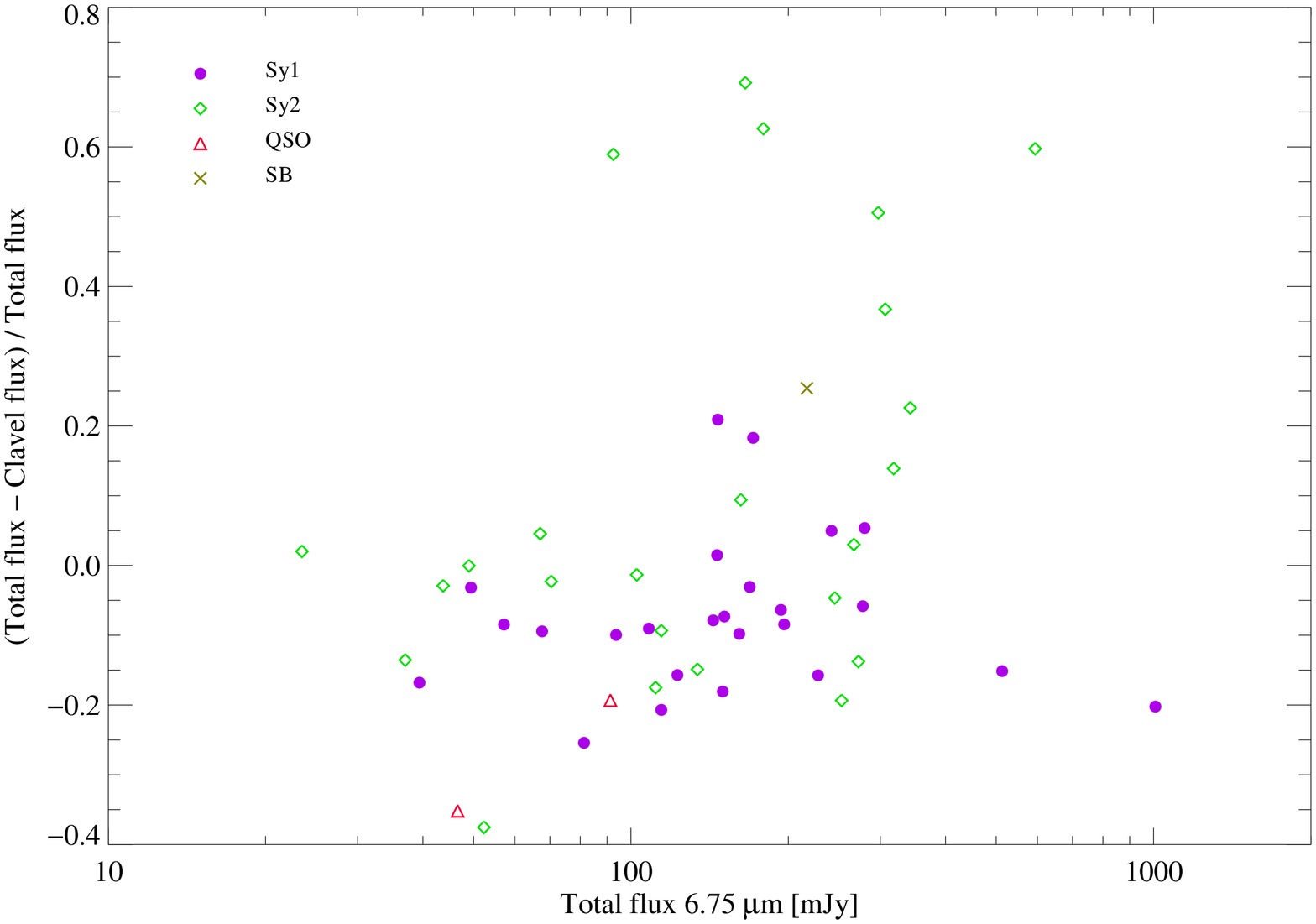}
\includegraphics[width=5.5cm,angle=90]{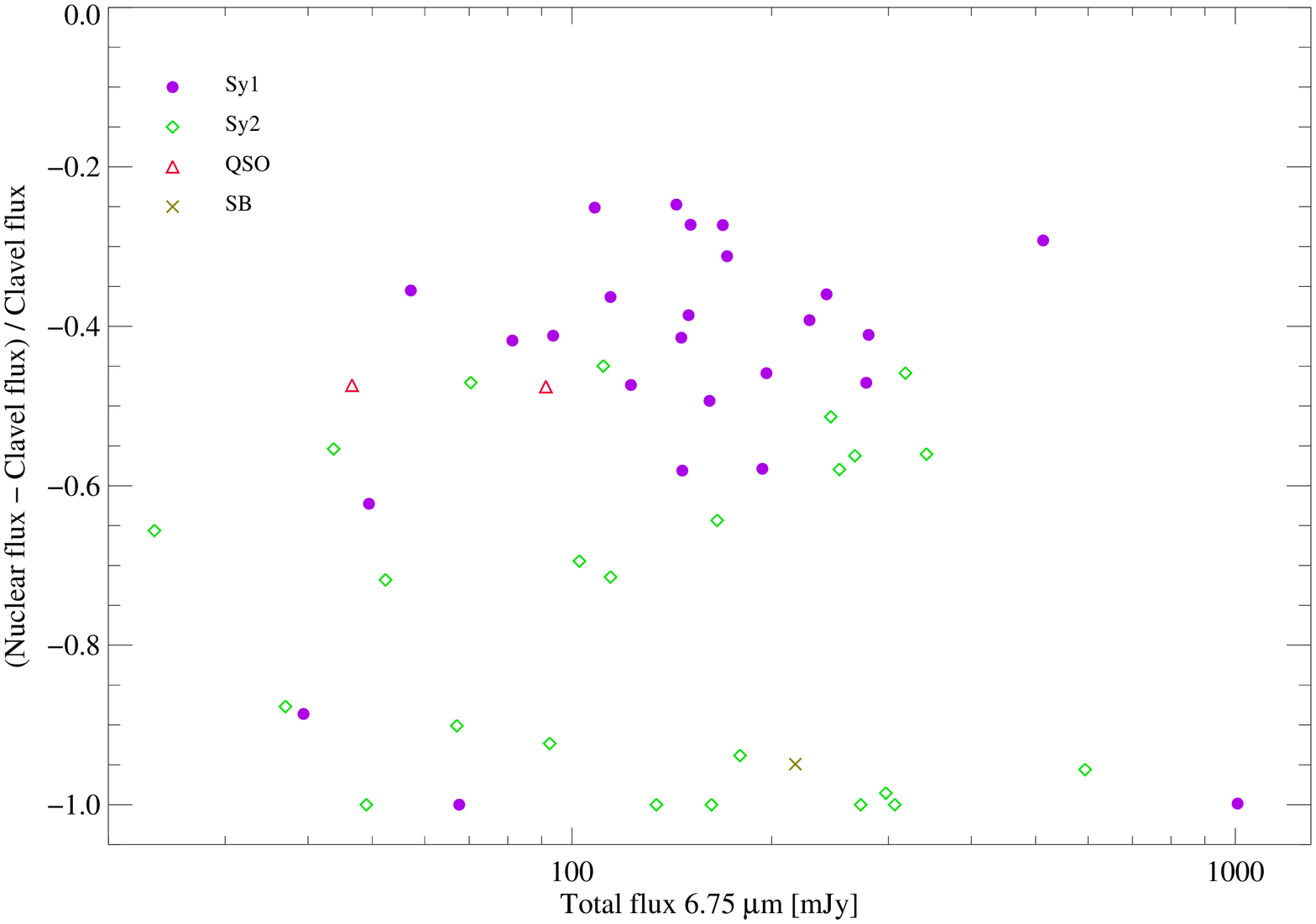}\par}
\figcaption{\footnotesize{Left panel: comparison of our total fluxes (see Table \ref{fluxes}) 
with the reported ones in \citet{Clavel00}. Filled circles represent Sy1 
(25 in total), open diamonds Sy2 (24 in total), open 
triangles QSOs (2 in total), and the starburst galaxy, NGC~701, is represented by 
a cross. NGC~1144 and Mrk~789 fluxes were not reported by C00. Right panel: the same comparison, but for 
our nuclear fluxes.}
\label{clavel1}}
\end{figure}

The main interest of our work is focused into a reliable
measurement of the nuclear flux, as well as its relative value to the total flux 
for each galaxy.
Once we have obtained the fluxes for each morphological component, 
reported in Table \ref{fluxes}, the first task we accomplished was 
the comparison of our results with the previous ones obtained from the same
data sample and reported by \citet{Clavel00} (hereafter C00). 
Here we recall that in C00, nuclear fluxes were obtained by integrating within a 
circular aperture of 3 pixel radius (9\arcsec) for point--like sources, 
and within an aperture of 4.5 pixels (13.\arcsec5) for extended ones.
In order to account for the 
emission in the wings of the PSF they also corrected by a factor 1.23. 
We note that this photometric correction makes sense for point sources, 
although their effect is not clear on extended sources like most of our 
targets are. We present in  Fig.~\ref{clavel1} the
difference between our total values and C00 measurements, normalized to our total 
fluxes. It can be seen that 
for most of the galaxies, our total fluxes are underestimated about 10\% compared 
with C00 ones. This might be due to the introduction of the correction for 
the PSF wings in C00 fluxes and not in our values, which is compensated by the fact that  
our measurements extended all over the detector, as opposite to C00, which uses a limited aperture. 
%Fluxes underestimated over 10\% correspond to brightness profiles with few points, 
%as is the case of QSOs. 
On the contrary, for large galaxies, as for example NGC~5033, NGC~5674, NGC~1241, 
and NGC~3982, our flux measurements are much higher than those reported by C00.
This discrepancy is due to the use of a 4.5 pixels aperture by C00 in clearly 
extended sources, resulting in underestimations of $\sim 60$\%.
We also present in Fig.~\ref{clavel1} the comparison between our nuclear fluxes 
and the C00 values, normalized to C00 values. 
In this case all our nuclear fluxes are lower than C00 ones, 
since we have subtracted the contribution of the galaxy to the nuclear flux, 
and moreover the 
radius of our Gaussian PSF is smaller than 3 pixels (the aperture employed by COO for point-like sources). 
Note that for Sy1 galaxies our values are in better agreement with C00, 
due to the fact that the compact nuclear component is dominant relative to the
total galaxy emission. 
Summarizing, we believe that the differences found between our flux measurements 
and those reported by C00 justify the need of 
a brightness profile decomposition in order to get a more accurate 
determination of the nuclear flux. 

%\clearpage

%\clearpage

%\begin{figure}[!h]
%\centering
%\includegraphics[width=8cm,angle=90]{f6_color.eps}
%\figcaption{\footnotesize{Same as Fig.~\ref{clavel1},
% but for our nuclear fluxes (see Table \ref{fluxes}).}
%\label{clavel2}}
%\end{figure}

\section{Results}

\subsection{Nuclear vs total emission}
\label{nuclear}

\begin{figure}[!h]
\centering
\includegraphics[width=9cm,angle=90]{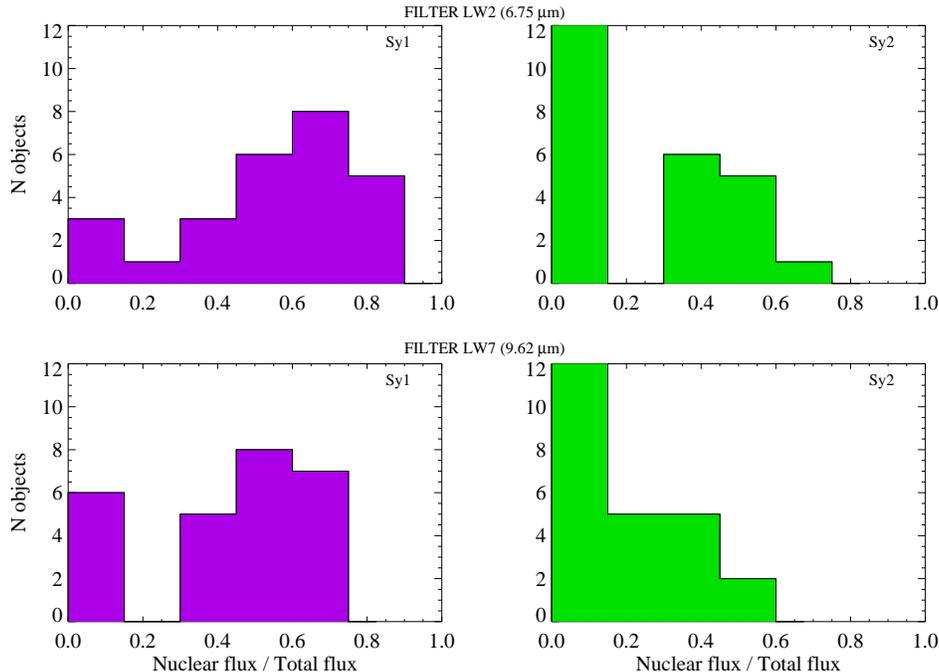}
\figcaption{\footnotesize{Histograms of nuclear vs total emission for Sy1 (left panels) and Sy2
galaxies (right panels) at 6.75 $\mu$m (top plots) and 9.62 $\mu$m (bottom plots). Note that
for Sy2 galaxies, most of them are concentrated around the lowest values of this ratio,
what does not happen in the case of Sy1.} 
\label{hist}}
\end{figure}

Once we have performed the profile decomposition we can measure 
the relative contribution of the nuclear source to the total emission for 54 objects of our sample. 
Note that for 7 galaxies our profile decomposition does not include a nuclear 
component (see Table \ref{params}), which includes 6 Sy2, namely NGC~1144, Mrk~3, 
NGC~1667, NGC~5728, and NGC~5953, plus  
the Sy1 Mrk~841 (it is an elliptical galaxy and the central component is  better fitted
with a bulge solely). 	
We show in Fig.~\ref{hist} 
the ratio of nuclear to total emission at 6.75 and 9.62 $\mu$m, respectively. 
It can be seen in both cases that there is a clear difference between the ratios 
for the Seyfert types 1 and 2. In fact, the
median of the ratio between nuclear and total fluxes for type 1 galaxies are 0.61 and 0.49, 
in the 6.75 and 9.62 $\mu$m filters, respectively. These values are both 0.14 for type 
2 galaxies. 
%For the QSOs in our sample the nuclear emission becomes dominant, in contrast with the 
%starburst galaxy, NGC~701, whose nuclear emission is quite small compared with the total flux.
%%\clearpage
%We have also represented the ratio of nuclear to total flux according to the Seyfert type and
%filter used but in the form of histograms in Fig.~\ref{hist}.  
In case of type 2 objects, most of them are grouped around the lowest values 
of the nuclear vs total flux ratio, whereas for type 1, the maximum number of
objects are located around or above the value 0.5. We have applied the Kolmogorov--Smirnov
test to check the significance level probability of the apparent difference between Sy1 and Sy2
distributions. We found that in both wavelenght ranges, the nuclear to total flux distributions 
for Sy1 and Sy2 are different in a 99.9\%.
%In the case of the LW2 filter, we obtain a normalized probability value of 3.66x10$^{-6}$, and 
%2.3x10$^{-4}$ for the distributions in the LW7 filter. These low values indicate that nuclear to total flux
%distributions for Sy1 and Sy2 in both filters are statistically different. 

We conclude that nuclear emission in the mid-infrared is a significative contribution of the total
flux in Sy1 galaxies, whereas for Sy2 other components overcome the
nuclear emission. This result is consistent with  the unification model predictions, since
the orientation of the molecular torus for type 1 Seyfert would be face-on with respect 
to our line of sight whereas for type 2 would be edge--on. 
Similar results have been found previously at different spectral ranges \citep{Yee83,Alonso96}.  

%\begin{figure}[!h]
%\centering
%\includegraphics[width=8cm,angle=90]{f7_color.eps}
%\figcaption{\footnotesize{Ratio between nuclear and total emission versus total emission at
%6.75 $\mu$m. Filled circles represent Sy1 (27 in total), open diamonds Sy2 (24
%in total), the two QSOs are represented by open triangles, and the starburst galaxy, NGC~701, by a cross.} 
%\label{psf1}}
%\end{figure}
%\begin{figure}[!h]
%\centering
%\includegraphics[width=8cm,angle=90]{f8_color.eps}
%\figcaption{\footnotesize{Same as Fig.~\ref{psf1}, but for fluxes measured at 9.62 $\mu$m.} 
%\label{psf2}}
%\end{figure}

%\clearpage

%\clearpage

\subsection{Nuclear and host galaxy mid-infrared colors}

\begin{figure}[!h]
\centering
\includegraphics[width=9cm,angle=90]{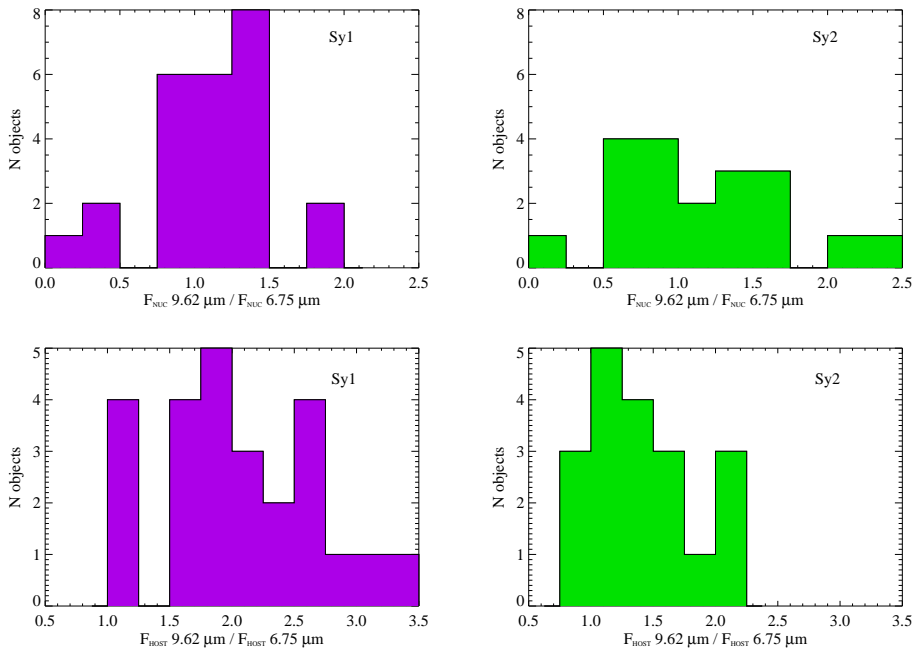}
\figcaption{\footnotesize{9.62/6.75~\micron\ colors for nuclear fluxes (top panels) and for
host galaxy fluxes (bottom). Left histograms represent Sy1 and right ones, Sy2.} 
\label{color1}}
\end{figure}

We have studied the mid-infrared color distributions for both the nuclear and
the host galaxy emission. The colors of host galaxies are computed after subtracting the
nuclear to the global emission. 
We represent in Fig.~\ref{color1}  the 9.62/6.75~\micron\  color histograms 
for both the nuclear and the host galaxy emission, for each Seyfert type.
The nuclear 9.62/6.75~\micron\ color distributions look very similar for both types, 
being the median values 1.14 and 1.02 for Sy1 and Sy2, respectively. 
We have performed a Kolmogorov-Smirnov test, obtaining a probability of 18\%, which 
indicates that both color distributions are not statistically different. 
Previous works \citep{Spinoglio89,Fadda98,Kuraszkiewicz03,
Alonso03,Lutz04,Rigby04} have found similar 
observed spectral energy distributions (SEDs) in the mid--infrared ($\lambda$ $>$ 5 \micron) for Sy1 
and Sy2 galaxies, 
These results are in conflict with the predictions of many compact torus models \citep{Pier92,Granato94,
Efstathiou95,Granato97}, that has promoted the search of a more distributed or complex geometry 
of the absorbing material around the AGN \citep{Nenkova02,Fritz06,Elitzur06,Ballantyne06}.

%\citet{Fritz06} predict no differences for type 1 and 2 in the region around 10~\micron\ when models
%with low optical depths and small outer-to-inner radii are considered.
Despite the small number of galaxies in our sample and the proximity of 
the bands analised, we can conclude that the nuclear mid-infrared emission 
seems to be  very similar and relatively flat in both types of Seyfert galaxies.
%In addition, we have also checked that the colors of the QSOs are in the middle of the 
%range of nuclear colors for Seyfert galaxies, whereas the color of
%the starburst NGC701 is among the bluest of the sample.

On the other hand, the color distributions of host galaxies look different, being those 
of type 1 redder than those of type 2 (median values of 2.00 and 1.41, respectively). 
The Kolmogorov-Smirnov test confirms that both distributions are different with a probability larger than 99\%.   
This result seems to contradict the core idea of Seyfert unification scheme, which 
predicts no much influence of the host galaxy to determine the type of active nucleus. 
In the same line of our result, \citet{Hunt99} pointed out that Sy1 could be older, more evolved 
than Sy2, since they are found more commonly in earlier morphological types. 

%\clearpage

%\clearpage

%\begin{figure}[!h]
%\centering
%\includegraphics[width=8cm,angle=90]{f11_color.eps}
%\figcaption{\footnotesize{Same as Fig.~\ref{color1}, but for nuclear fluxes.} 
%\label{color2}}
%\end{figure}

%according with
%the results obtained by \citet{Fritz06} for a modelling of the emission by dust in a torus geometry, 
%a flared disk, and a dust grain distribution function including a full range of grain sizes. This model
%predicts redder  9.62/6.75~\micron\  colors for Sy1 than for Sy2, specially for the lowest equatorial optical
%depths and outer-to-inner radii ratios. 

\subsection{The radio/mid--infrared correlation in our sample}
\label{radio}

A well-known correlation between global far--infrared and  radio emission from galaxies
apply to a wide range of Hubble types \citep{Fitt88,Hummel88,Wunderlich87}. 
%This correlation is displayed over 3 orders 
%of magnitude in luminosity, from dust--rich dwarfs to ultraluminous infrared galaxies (ULIRGs). 
The most natural explanation for such correlation is related to 
star--formation activity.
In addition, \citet{Elbaz02,Gruppioni03} found that it
can be also extended to the mid--infrared range although with larger dispersion. 
%than in the case of the radio/far-infrared. 
%A priori, there is not obvious reason why such correlation should exist. The mid-infrared emission 
%is mainly produced by 
%a mixture of complex PAH plus thermal emission from small dust grains at 
%high temperatures, whereas the far--infrared emission is dominated by large dust grains at lower
%temperatures. 
%More recently, \citet{Appleton04} presented  evidences that the radio/far- and mid--infrared
%correlation is valid to cosmologically significant distances (up to $z\simeq1$).
Thus, the study of the radio/far--infrared correlation in active galactic nuclei turns out to be
an useful tool  for studying the starburst--AGN relationship.
A priori, the presence of nuclear radio activity not related to supernova remnants
should introduce departures from the mentioned correlation. 
\citet{Roy98} reported that Seyfert galaxies 
also display the radio/far-infrared correlation, although with a larger scatter than
non-active galaxies. They also found that Seyfert with compact radio cores tend to 
deviate from the correlation, contrary to those without compact cores. However, they
noted that the correlation does not improve after subtraction of the compact radio emission.

\begin{figure}[!h]
\centering
\includegraphics[width=8cm,angle=90]{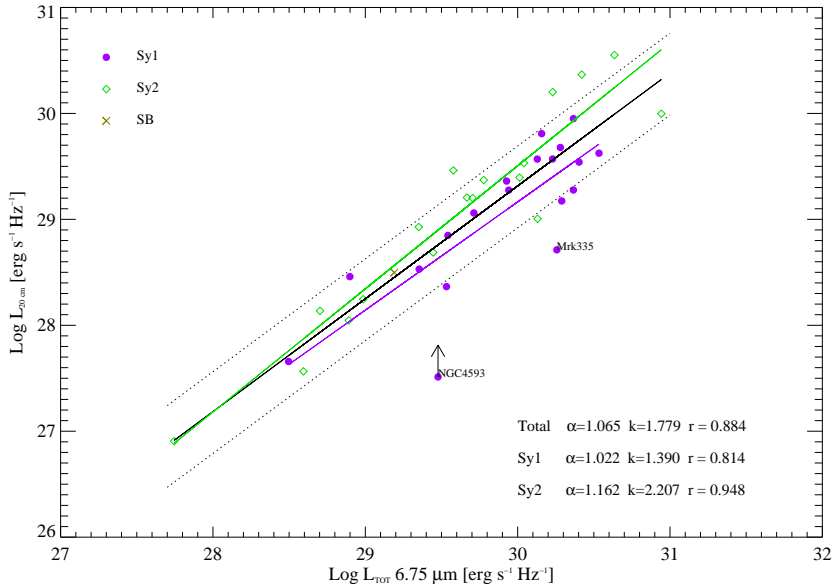}
\figcaption{\footnotesize{Radio luminosity (L$_{20cm}$) versus mid-infrared global luminosity 
(L$_{TOT}$~6.75~$\mu$m). Filled circles represent Seyfert 1 (19 in total), 
open diamonds Seyfert 2 (18 in total) and the cross the starburst galaxy, NGC~701. 
The continuous black line represents a linear fit to all data, 
the dashed lines correspond to the limits of the correlation at one $\sigma$.  
The values of $\alpha$ and k are derived from the expression 
$L_{20cm}/10^{28}=k\cdot(L_{TOT}/10^{29})^{\alpha}$ and are given explicitely in the 
legend. Separate fits to types 1 and 2 data are also presented.}
\label{rush1}}
\end{figure}

We have investigated the presence of the radio/mid-infrared correlation in our
sample of Seyfert galaxies.  The radio data were taken from those available in 
the literature. Most of them (data of 32 galaxies) come 
from the sample observed by \citet{Rush96} at 20 cm using the VLA\footnote{For another 6 objects 
(Mrk~789, NGC~4579, ESO141-G55, NGC~4593, Mrk~509, and NGC~701)
we have used VLA measurements at 20~cm, reported by different authors
\citep{Ulvestad84,Condon98,Ho01,Wadadekar04}.}.
The resolution ($\sim$ 1.\arcmin5) of these radio data samples the emission of large scales 
in the galaxy, namely $\sim$ 35~kpc at the median distance of the sample.  
%These structures should be similar to those reported by \citet{Baum93}.
From here and in all later sections, we will present results only using the LW2 filter, 
but we have always checked that the results obtained 
with both filters are practically the same. 
% We present in Fig.~\ref{flujo_total} the radio  and 6.75~\micron\ total flux--flux scatter diagram. 
We have excluded two galaxies from subsequent analysis because their behaviour deviates from
the rest of the sample. They are identified as NGC~4593 (its radio flux is a lower limit) 
and Mrk~335 (a S0 galaxy with a low radio emission likely due to be an early type galaxy).
%As can be seen there is an apparent correlation between the 20~cm and 6.75~\micron\ data sets.
The luminosity-luminosity scatter diagram is presented in Fig. \ref{rush1}. A linear correlation
appears in log-log scale, with a slope of 1.07 and a coefficient $r= 0.88$\footnote{In order 
to verify that the correlation is not a distance effect, we have performed a test 
by upsetting in a random way the redshifts ($z$) of galaxies. If the linear correlation 
is merely a distance effect, it should be mantained with any $z$ distribution. In our case,
the correlation disappears when the distance distribution is changed, confirming that 
the observed correlation is not a distance effect.}.  
Similar correlations are found when the data are fitted for types 1 and 2, separately 
(see Fig. \ref{rush1}). The highest correlation coefficient ($r=0.95$) appears when only 
Sy2 are considered. 
In addition, we have applied a non--parametric correlation 
test, such as the Spearman's rank correlation test. The results indicate that 
the correlation is significant ($p <0.01$) in all cases (all galaxies, and Sy1 and Sy2, 
separately). 

%\clearpage

%\clearpage

\begin{figure}[!ht]
\centering
\includegraphics[width=8cm,angle=90]{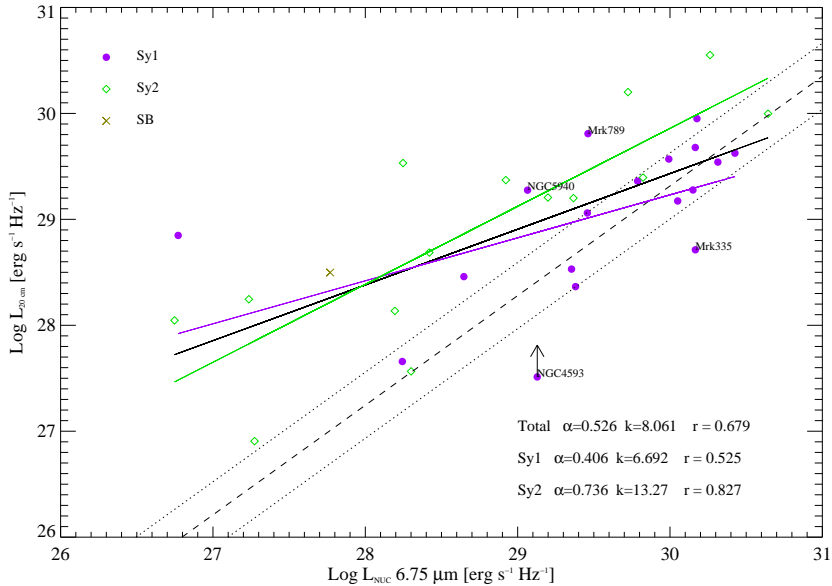}
\figcaption{\footnotesize{Radio luminosity (L$_{20cm}$) versus mid-infrared nuclear 
luminosity (L$_{NUC}$~6.75~$\mu$m). The meaning of symbols is the same as in 
Fig.~\ref{rush1}. The continuous line represents the fit to all galaxies, 
the dashed line represents the correlation using global mid-IR luminosity, 
from Fig.~\ref{rush1}. 
The corresponding fits to Sy1 and Sy2 data separately, are also represented.}
\label{rush2}}
\end{figure}

We have also explored the existence of such a correlation between the same radio data and 
our estimations of the nuclear mid-infrared emission. The luminosity scatter diagram
is presented in Fig. \ref{rush2}. 
In this case, the correlation appears worse ($r \simeq 0.68$) than when global 
mid-infrared luminosity is considered. Moreover, the correlation becomes distant
from linear in log-log space, being the slopes different from unity (see values 
in Fig. \ref{rush2}). We also noticed that the correlation slightly improves when
only Sy2 galaxies are considered.
The less luminous galaxies (mostly Sy2) display an excess of radio luminosity compared 
to their nuclear mid-IR luminosity, indicating that most of the radio emission
in these objects is not related to nuclear processes. 
Summarizing, the global radio emission seems to be 
related closely to AGN activity in the most nuclear-infrared luminous galaxies (mostly Sy1), 
whereas for the less luminous the radio emission would be more related to non-nuclear 
stellar processes. In this respect it is worth to mention the result of
\citet{Baum93}, who found that extra-nuclear (several kpc) radio emission, similar
to the lobes of powerful radio galaxies, appears frequently in galaxies whose properties
are dominated either by an AGN or a starburst.

%\clearpage

%\clearpage

According to our results, we claim that for the less radio luminous galaxies (mostly Sy2), 
radio emission is more related to stellar processes, given the
fact that most of the mid-infrared emission does not come from the nuclear region. 
However, for the more luminous galaxies (mostly Sy1),the correlation could be 
attributed to extended radio emission which is somewhat related to 
the presence of the AGN as claimed by \citet{Baum93} and indirectly by \citet{Roy98}.

\subsection{Comparison among X-rays and mid-infrared emission}

	The hard X--ray (2-10 keV) spectral region is of particular interest for 
the study of AGN.  It provides a direct view to the central engine and it is believed to 
be a reasonable isotropic indicator of the bolometric luminosity of the
nuclei. On contrast, in the simple unified models, the mid-infrared emission is expected to vary 
as a function, not only of the AGN luminosity, but also of the distribution of 
the obscuring matter along the viewing direction of the observer. 
As a result of this, the comparison of the mid-infrared versus 
the hard X--ray measurements constitutes an important test for unification models. 
% Previously, a tight correlation between the two quantities had been reported by \citet{Krabbe01}, 
% using a very reduced sample of only 8 Seyfert galaxies. 
% Later, \citet{Lutz04} found a similar correlation for a larger sample of 71 Seyferts. 
% However, they found considerable scatter in the ratio between intrinsic (absorption corrected) 
% X--ray and mid-infrared emissions, being higher for type 2 Seyferts. They ascribe part of this dispersion  
% to variations in the AGN spectral energy distribution and geometry of obscuring dust, as well as 
% time variability. 
% More surprisingly, \citet{Lutz04} reported no significant difference between type 1 and 2 Seyferts 
% in the average ratio of X-ray and mid-infrared continuum. They explained this fact as a consequence of
% a large isotropic contribution from extended dust components which largely dilutes 
% the anisotropic emission from the torus. 
% In a recent work, \citet{Horst06} claimed a tight correlation between 
% nuclear mid-infrared  and intrinsic hard X--ray luminosities, using 
% very high spatial resolution mid-infrared images from VISIR on the VLT. They found the best fit power law index
% different of unity ($log L_{MIR} \propto 1.6~log L_X$). 

%Smooth torus models predict a flux difference between type 1 and 2 nuclei of
%an order of magnitude. However, clumpy torus models \citep{Nenkova02, Honig06} 
%are compatible with {\bf no flux difference in the mid-infrared}. Clumpy torus can appear 
%as optically thin in the mid-infrared, although individual clouds appear optically thick. 
 
\begin{figure}[!h]
\centering
\includegraphics[width=8cm,angle=90]{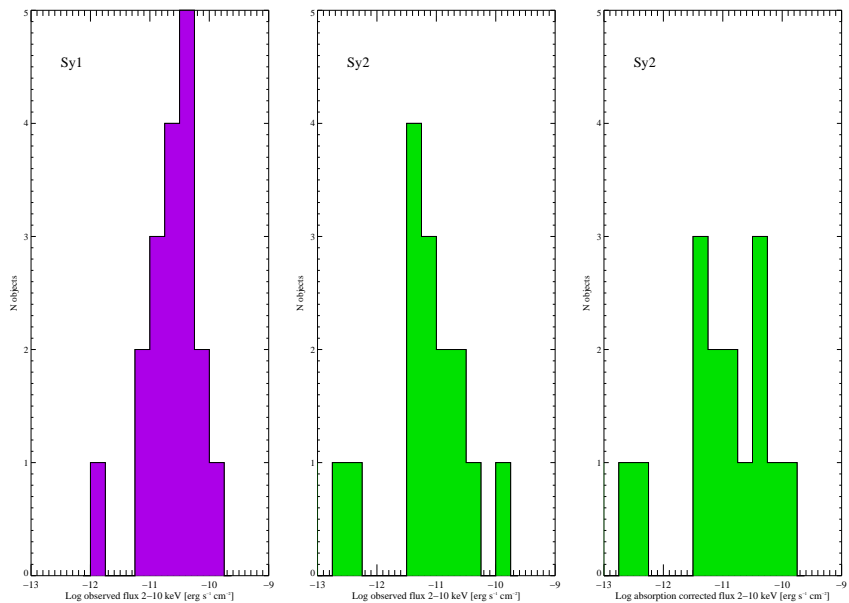}
\figcaption{\footnotesize{Histograms of observed hard X--ray fluxes for Sy1 (left), Sy2 (center), and 
absorption corrected hard X--ray fluxes for Sy2 (right).} 
\label{lutz_hist}}
\end{figure}

	Here, we compare our mid-infrared measurements with the hard X-ray fluxes compiled by 
\citet{Lutz04}, obtained from various literature sources, taken by different satellites 
(e.g. ASCA, BeppoSAX, Chandra, XMM-Newton). 
They reported measurements of  35  objects which are also included in our sample, 
from which only 27 galaxies have intrinsic or absorption--corrected values. 
%X-ray emission is significantly more absorbed in Sy2 than in Sy1, then, for Sy2 for which 
%there are absorbing column density N$_{H}$ measurements, 
%\citet{Lutz04} report the intrinsec absorption-corrected hard X-ray fluxes (unfortunately,
%there are only 27 galaxies in our sample with intrinsec fluxes measurements). 
%We have compared these data with our nuclear and total MIR fluxes. 
%, in order to verify the conclusions found by \citet{Lutz04}, who calculated their 6\micron fluxes
%by integration of continuum from ISOPHOT spectra. 
We represent in Fig. \ref{lutz_hist} the histograms corresponding to 
observed hard X--ray fluxes for Sy1 and Sy2, plus absorption corrected hard X--ray fluxes for Sy2.
%We have then performed the analysis of this section based only in the filter centered at 6.75~\micron, 
%after having verified that there are not significative differences between both ranges. 
% There is a clear segregation between the two types
% of Seyfert galaxies, being type 2 nuclei less luminous in X-rays (and, as we have seen before, in 
% mid-infrared). 
Given the fact that the hard X--ray emission in Sy2 galaxies is commonly strongly absorbed, 
it makes sense to consider the intrinsic, instead than the observed, hard X--ray fluxes. 
For those cases in which absorption corrected fluxes are not provided we adopt the observed
flux as a lower limit. 
%Sy2 nuclei show similar hard X--ray intrinsic fluxes,  
% and generally lower than those of type 1. 
%The mid-infrared fluxes of Sy2 nuclei are 
%much lower than those of type 1. 
In this section we do not refer to the total infrared flux since hard X--ray emission is
expected to be uniquely related to nuclear emission. 
%Indeed, there is not  
%an apparent correlation between X--ray and total mid-infrared luminosities 
%\textcolor{red}{Hemos comprobado esto?}. 

%It is worth to mention that the value of the X-ray to mid-infrared ratio for the QSO
%H 1821+643 is equal to -1.42, which is among the type 1 nuclei values.

%\clearpage

\begin{figure}[!h]
\centering
\includegraphics[width=8cm,angle=90]{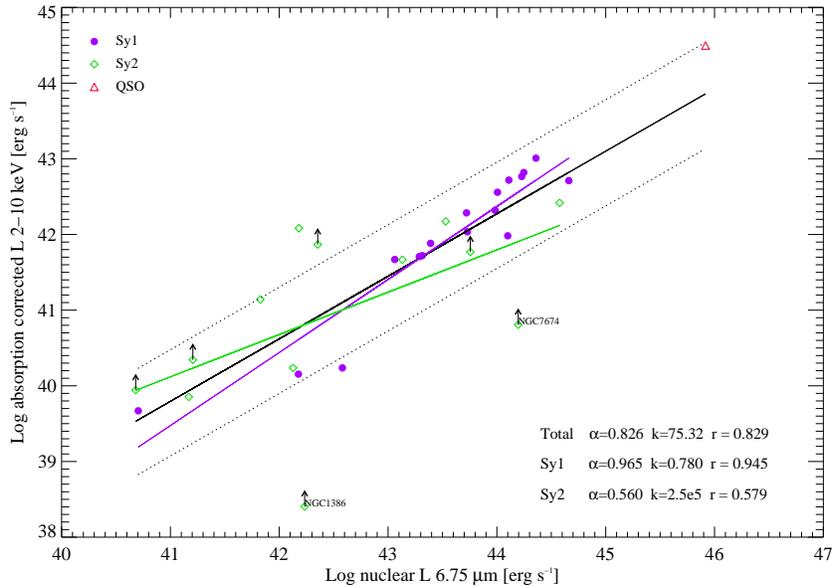}
\figcaption{\footnotesize{Absorption corrected hard X-ray luminosities versus nuclear
6.75~\micron\ luminosities. 
Continuous line in black is the total fit and dotted lines correspond to one $\sigma$ limits. 
Filled circles represent Sy1 (17 in total), open diamonds Sy2 
(13 in total), and the QSO H~1821+643 is represented by an open triangle. 
Fits to Sy1 and Sy2 data are represented too.
Observed hard X--ray luminosities are represented as lower limits when no N$_{H}$ data is reported.} 
\label{lutz_nuc_lumin}}
\end{figure}

We have also looked for a correlation between X--ray and mid-infrared luminosities
(see Fig.~\ref{lutz_nuc_lumin}).
The data corresponding to all galaxies display a linear correlation
of slope 0.8 with correlation coefficient r = 0.83.
We noticed that a good correlation appears between X--ray and nuclear 
mid-infrared luminosities for type 1 nuclei ($\alpha$ = 0.97 and r = 0.95). However, 
the data for type 2 nuclei present higher dispersion 
($\alpha$ = 0.56 and r = 0.58), showing lower mid-infrared luminosities relative to their X--ray luminosities.
We have also applied the Spearman's rank correlation test, 
finding that the correlation is significant when all galaxies and only type 1 nuclei 
are considered. 
%However, in the flux-flux space diagram, we found not correlation, so we are in the
%same situation as we see in Section \ref{radio} for the nuclear emission: if the intrinsic correlation between X-ray and
%mid-infrared luminosities is not linear ($\alpha \neq 1$), then a random redshift distribution will smear out any
%flux-flux correlation \citep{Feigelson83}.
We have checked that this correlation is not a distance effect by upsetting in a random way the 
redshift distribution of the galaxies.

The ratio of hard--X ray to nuclear mid-infrared emission 
appears larger in the case of type 1 nuclei 
($<log (L^X_{intr}/L^{MIR})>=-1.62\pm0.35$), than in the case of type 2
nuclei ($<log (L^X_{intr}/L^{MIR})> -1.19\pm0.67$).  Both distributions appear
significatively different according to the Kolmogorov-Smirnov test.
We have excluded the galaxies NGC1386 and NGC7674 due to their extremely low ratios, 
which in addition are lower limits.

The difference in the ratios for Sy1 and Sy2 appears contradictory to that has been claimed by \citet{Lutz04}
and \citet{Horst06}. Nevertheless, we expect to find different results when we compare with the 
work of \citet{Lutz04}, since they have extracted the nuclear emission by spectral 
decomposition of very large aperture data. We have checked that their estimations
of the AGN contribution are largely overestimated compared to our measurements
of  the nuclear component. In the case of \citet{Horst06}, they have obtained
very high resolution data in the mid-infrared range but their sample is very reduced. 
The small dispersion claimed by \citet{Horst06} could be due to the limited 
number of galaxies included.

Our results are in qualitative agreement with the predictions of unification models. 
In case of type 1 nuclei we would see direct emission from the nucleus at any wavelength
range. In type 2 nuclei  the intrinsic hard X--ray 
emission would be similar, {\it i.e.}, after absorption correction, to type 1. 
The mid-infrared emission coming from a dusty torus would depend on the viewing angle as a 
function of the optical depth of the obscuring structure. The predictions vary 
drastically for different models. The initially proposed torus models by \citet{Pier92} predict
large variation for both Seyfert type in the mid-infrared range. However, more recent clumpy 
torus models do not predict large offset between both Seyfert types \citep{Nenkova02, Honig06}.
This variation will be larger for shorter wavelengths where the innermost part of the
torus dominates, and will be attenuated for longer wavelengths where the outermost
parts dominate. This may be one of the reasons to explain why we detect large difference
in the X--ray to mid-infrared ratio for types 1 and 2, contrary to that has been found by previous
authors using a longer wavelength filter \citep{Horst06}. 

%\clearpage
   
%\clearpage

% Summarizing, we have found a general trend for Sy1 to show larger observed X--ray and mid-infrared nuclear fluxes 
% as compared to Sy2. The intrinsic (absorption corrected) X--ray fluxes are very similar for both types of nuclei, which 
% almost washed out the correlation between X--ray and mid-infrared fluxes. The luminosity--luminosity scatter diagram
% seems to indicate a good correlation between the two quantities, at least in the case of type 1 nuclei.

\subsection{The mid-infrared properties of hidden broad-line region Seyfert 2 and narrow-line Seyfert 1 galaxies}

According to the unification scheme of Seyfert galaxies each Sy2 galaxy should have a hidden
broad--line region (HBLR), as found in the prototypical NGC~1068 \citep{Antonucci85}. 
However, spectropolarimetric surveys of complete samples of Sy2 galaxies show that hidden type 1 
nuclei have been detected in less than $50$~\% of the  galaxies from the CfA and 12~\micron~samples 
\citep{Lumsden01, Moran00, Tran01,Tran03}. 
These objects with failed detection of a type 1 nuclei are known as non--hidden broad-line 
region (NHBLR) Sy2 galaxies. 
The non--detection of broad lines could be explained in the case of an edge-on line of sight
or in the absence of an electron scattering region \citep{Miller90,Taniguchi99}.
However, some large--scale characteristics of the HBLR galaxy population are not shared by
the non--HBLR population.   
The HBLR galaxies display distinctly higher radio power relative to their far-infrared output and 
hotter dust temperature (F$_{25\micron}/F_{60\micron}$ color), 
compared to the NHBLR Sy2 galaxies \citep{Tran03}. 
The NHBLR galaxies also appear sistematicly as weaker radio sources
than their HBLR counterparts \citep{Thean01}. 
The level of obscuration, as measured by the Balmer decrement, is indistinguishable between both types
of Sy2 as well as the high level of starlight domination \citep{Moran00}.
Thus, the relative number  of HBLR and non--HBLR galaxies 
cannot be explained by different orientations, 
challenging the unification scheme \citep{Tran01,Tran03,Lumsden01}.
These results strongly support the existence of two intrinsically different populations of Sy2
galaxies: one harbouring an energetic, hidden Sy1 nucleus with a broad-line region 
and the other, "true" Sy2 galaxies, with a weak or absent type 1 nucleus and a strong, 
likely dominating starburst component \citep{Tran03}. 

\begin{figure*}
\centering
\includegraphics[width=8cm,angle=90]{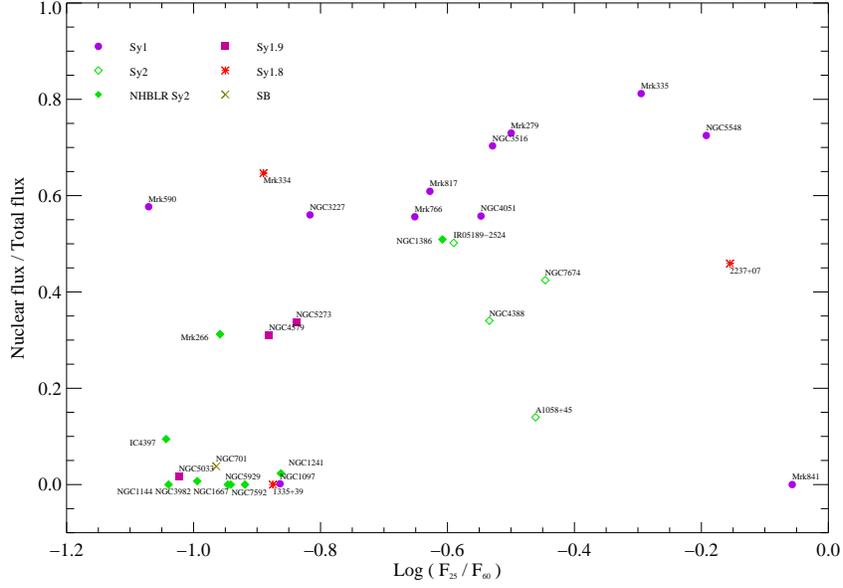}
\figcaption{\footnotesize{Nuclear/total mid-infrared flux ratio versus the far-infrared 
color $F_{25\micron}/F_{60\micron}$
for galaxies in our sample. NHBLR are shown as filled diamonds and confirmed HBLR 
as open diamonds. Circles represent Sy1, squares Sy1.9, 
asterisks Sy1.8, and the starburst galaxy is marked with a cross. IC~4397 and NGC~7592 are represented  
as NHBLR type 2 nuclei based only in their position in this diagram.} 
\label{tran1}}
\end{figure*}

Three galaxies from \citet{Tran03}
with spectropolarimetric confirmed HBLR are included in our initial sample of 57 AGNs, 
namely NGC~4388, NGC~7674, and IR~05189-2524, and also
seven NHBLR, namely Mrk~266, NGC~1144, NGC~1241, NGC~1386, NGC~1667, NGC~3982, and NGC~5929. 
We have investigated possible differences in the mid--infrared properties segregating both types of 
Sy2 galaxies in our sample. We add two Sy2 galaxies to the \citet{Tran03} ones, namely IC~4397 and NGC~7592, and other
Seyfert types for which we have found far-infrared data in the literature \citep{Perez01}.
We present in Fig. \ref{tran1} the variation of the ratio between our mid-infrared nuclear and total
fluxes versus the far--infrared color $F_{25\micron}/F_{60\micron}$.  
The nuclear versus total emission ratios of the HBLR are among the highest values for Sy2. 
On the contrary, for NHBLR these ratios are among the lowest values. 
The ratio of the nuclear versus total flux is below 0.02 in NGC~1241 and NGC~3982, and does not exceed
0.5 in Mrk~266, IC~4397, and NGC~1386. For the  galaxies NGC~1144 and NGC~1667 we could not identify 
a central unresolved source, which could be associated with the presence of an AGN. 
NGC~5929 and NGC~7592 are excluded from this analysis  because their nuclear emission cannot be isolated 
because of its close--by companion, and because of its double nucleus, respectively, 
due to our limited spatial resolution. Both galaxies appear in Fig. \ref{tran1} in order to show their 
far--infrared colors.

%\clearpage

%\clearpage

\begin{figure*}
\centering
\includegraphics[width=8cm,angle=90]{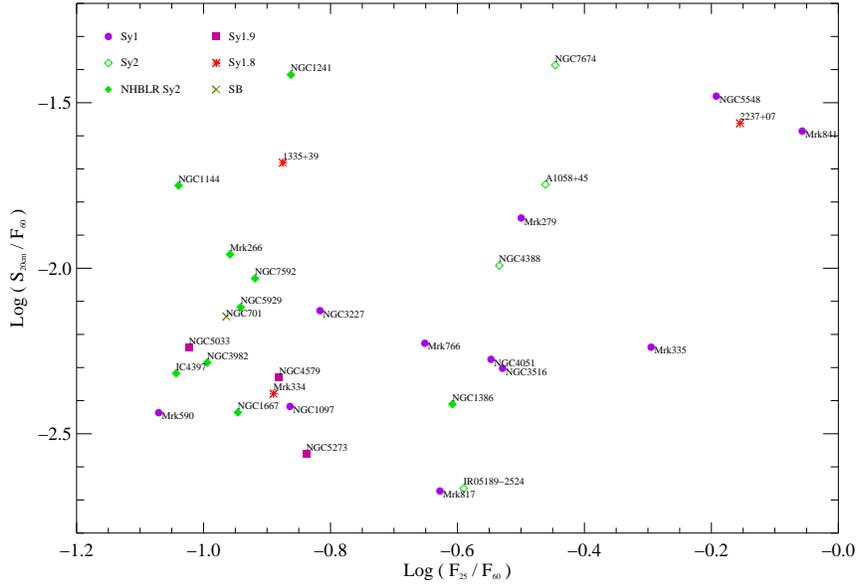}
\figcaption{\footnotesize{Ratio of radio flux density S$_{20cm}$ and far-infrared flux 
F$_{60\micron}$ as a function of
F$_{25\micron}/F_{60\micron}$ for galaxies of our sample. IC~4397 and NGC~7592 are represented 
as NHBLR type 2 nuclei based only in their position in this diagram. Symbols are the same as in Fig.
\ref{tran1}} 
\label{fir}}
\end{figure*}

\citet{Tran03} found that the far--infrared colors $F_{25\micron}/F_{60\micron}$ compared to the radio
flux can be used as a good discriminant between HBLR and NHBLR (Fig. 4 in Tran 2003). 
We present a similar diagram in Fig. \ref{fir} for the galaxies in our sample.
Radio fluxes (20 cm) are the same used in Section \ref{radio}. 
Despite the low number of objects, it appears that recognized HBLR Sy2 galaxies are located
in the upper--right corner, whereas NHBLR galaxies tend to occupy the bottom--left corner of the diagram. 
As a result of the use of Figs. \ref{tran1} and \ref{fir} as diagnostic diagrams we can propose new NHBLR 
candidates. For instance, IC~4397 and NGC~7592 are located in the left--bottom quadrant 
of Fig. \ref{fir}, and the former also shows a low nuclear to total flux ratio. 
These facts would support the classification of IC~4397 and NGC~7592 as possible NHBLR Sy2 nucleus.   

%\clearpage

%\clearpage

It is interesting to look at the position of intermediate Seyfert type galaxies in Fig. \ref{fir}. 
\citet{Tran01} only reports pure Sy2 data, but in Fig. \ref{fir} we are including
Sy1.8 and Sy1.9 too. All the three Sy1.9 have low S$_{20cm}$/F$_{60\micron}$ and 
F$_{25\micron}/F_{60\micron}$ ratios, corresponding to
the NHBLR region. Sy1.8 galaxies ocuppy random positions in the diagram.

\citet{Zhang06} suggest that NHBLR Sy2 are the counterparts of Narrow Line Seyfert 1 (NL Sy1)
viewed at larger angles.
The NL Sy1 class is characterized for very narrow Balmer lines 
[H$_{\beta}$ FWHM $\le$~2000 km~s$^{-1}$],
strong [Fe II] lines \citep{Osterbrock85}, and violent variability in soft X-rays \citep{Boller96}. 
%They likely contain less massive black holes at 
%the Eddington limit rates \citep{Boller96,Laor97} and can be described by a slim disk \citep{Wang99,Chen04}. 
We have looked for galaxies in our sample classified by them as NL Sy1 and we found three: 
Mrk~335, Mrk~766, and NGC~4051, whose fluxes are reported in Table \ref{fluxes}. In our diagrams 
(Figs. \ref{tran1} and \ref{fir}) these galaxies occupy the same 
region as Sy1 and HBLR Sy2. According to our view this result contradicts the hypothesis
of \citet{Zhang06}. 

\section{Conclusions}

	We have presented and analyzed mid-infrared data of a sample of Seyfert galaxies obtained with the 
instrument ISOCAM, being able to separate the nuclear and extended contributions to the total emission. 
The following results were found:

\begin{itemize}
\item The nuclear emission in the mid-infrared is a significant contribution to the total flux in Sy1 galaxies
whereas for Sy2 other components overcome the nuclear emission. 
This result is consistent with the unification model predictions. 
\item The mid-infrared color distribution of the host galaxies of Sy1 appears redder than that of Sy2, 
whereas the nuclear mid-infrared emission seems to be more similar in both types of Seyfert galaxies, being 
also relatively flat. 
% The colors of the QSOs are in the middle of the range of nuclear colors for Seyfert galaxies, 
% whereas the color of the starburst is among the bluest of the sample.

\item The global radio emission of Seyfert galaxies seems to be 
related closely to AGN activity in the most nuclear-infrared luminous galaxies (mostly Sy1), 
whereas for the less luminous (mostly Sy2) the radio emission would be more related to non-nuclear 
stellar processes.

%\item Seyfert galaxies in our sample displayed the radio/mid-infrared correlation, when global 
%mid-infrared emission is considered, although appearing less tight that those
%displayed by non-active spirals and starburst galaxies. This is naturally attributed to a relevant
%contribution of non-stellar process from the AGN.
%When only the nuclear mid-infrared emission is considered for the Seyfert sample, 
%the radio emission seems to be related closely to the nuclear mid-infrared source in the most
%luminous objects, whereas for the less luminous ones there seems to be an excess of radio emission.

\item The luminosity--luminosity scatter diagram between hard X-ray and mid-infrared emission seems to indicate 
a good correlation between the two quantities, at least in the case of type 1 nuclei.
The ratio between the intrinsic hard X--ray and the 
nuclear mid-infrared emission presents large scatter and slightly larger values for type 2 
Seyfert galaxies. These results seem to be consistent with the presence of a clumpy 
dusty torus surrounding the active nucleus.% scatter diagram seems to indicate a good correlation between the two quantities, at least in the case 
% of type 1 nuclei.
% A general trend is found for Sy1 to show larger observed X--ray and mid-infrared nuclear fluxes 
% as compared to Sy2. The intrinsic (absorption corrected) X--ray fluxes are very similar for both types of nuclei, which 
% almost washed out the correlation between X--ray and mid-infrared fluxes. The luminosity--luminosity 
% scatter diagram seems to indicate a good correlation between the two quantities, at least in the case 
% of type 1 nuclei.
\item The mid-infrared properties of HBLR and NHBLR nuclei appear different. The nuclear to total flux
ratios of the HBLR objects are among the highest values for Sy2. On the contrary, for NHBLR these ratios
are among the lowest values. A diagram representing the mid-infrared 
nuclear to total emission ratio versus the far-infrared colors seems to be a useful tool to segregate both
types of Sy2 nuclei.
\end{itemize}

\acknowledgments

We thank the anonymous referee for his very useful comments that led to the improvement of our work.
This research has made use of the NASA/IPAC Extragalactic Database (NED) which is operated by the 
Jet Propulsion Laboratory, California Institute of Technology, under contract with the National 
Aeronautics and Space Administration.
This publication makes use of data products from the Two Micron All Sky Survey, which is a joint 
project of the University of Massachusetts and the Infrared Processing and Analysis 
Center/California Institute of Technology, funded by the National Aeronautics and Space 
Administration and the National Science Foundation. We thank the anonymous referee for his very useful comments that led to the improvement of our work.

\clearpage

\begin{table}
\scriptsize
\centering
\begin{tabular}{lccccc}
\hline
\hline
TARGET&Morphology&Type&$z$&Exptime (s)& Start Time (UT)\\
\hline
NGC1097	     &RSB(r)b	     & Sy1		 & 0.00425 	 &516   &  01 01 1997 02:20:48	\\
NGC1125	     &SAB0	     & Sy2		 & 0.01100	 &516   &  01 02 1998 13:42:51	\\
NGC1144	     &RSApec	     & Sy2		 & 0.02885 	 &516   &  13 07 1997 05:44:46	\\
NGC1241	     &SB(rs)b	     & Sy2		 & 0.01351	 &516   &  04 01 1998 15:03:35	\\
Z~1335.5+3925& S?	     & Sy1.8		 & 0.02009	 &516   &  09 06 1996 12:45:19	\\
NGC1386	     & SB(s)0+	     & Sy2		 & 0.00289	 &516   &  27 01 1998 02:50:27	\\
NGC1566	     & RSAB(rs)bc    & Sy1		 & 0.00499	 &516   &  18 05 1997 18:08:52	\\
NGC1667	     & SAB(r)c	     & Sy2		 & 0.01517	 &516   &  02 10 1997 21:32:21	\\
Mrk266	     &pec	     & Sy2/SB		 & 0.02786	 &516   &  04 05 1996 16:04:24	\\
Mrk279	     &S0	     & Sy1.5		 & 0.02940 	 &514   &  05 02 1996 07:26:57	\\
NGC3227	     &SAB(s)pec      & Sy1.5		 & 0.00386	 &516   &  25 04 1996 04:32:37	\\
Mrk334	     &pec/HII	     & Sy1.8		 & 0.02196	 &516   &  12 12 1996 00:59:23	\\
Mrk335	     &S0/a	     & Sy1.2		 & 0.02564	 &516   &  12 12 1996 01:29:13	\\
NGC3516	     &RSB(s)0	     & Sy1.5		 & 0.00884	 &524   &  12 03 1996 11:32:22	\\
NGC3982	     &SAB(r)b	     & Sy2		 & 0.00370	 &524   &  08 04 1996 06:14:36	\\
3C382	     &BLRG	     & Sy1		 & 0.05787	 &514   &  16 02 1996 18:30:26	\\
Mrk3	     &S0:	     & Sy2		 & 0.01351	 &516   &  04 09 1997 11:12:01	\\
NGC4051	     &SAB(rs)bc      & Sy1.5		 & 0.00242 	 &516   &  09 05 1996 09:36:57	\\
IC4329A	     & SA0+	     & Sy1.2	 	 & 0.01605 	 &514   &  14 02 1996 16:25:00	\\
NGC4388	     &SA(s)b	     & Sy2	 	 & 0.00842	 &516   &  09 07 1996 06:18:36  \\
IC4397	     & S?	     & Sy2/HII  	 & 0.01474	 &514   &  07 02 1996 19:08:30	\\
NGC4507	     & SAB(s)ab	     & Sy2		 & 0.01180	 &514   &  04 02 1996 12:06:42	\\
NGC4579	     & SAB(rs)b	     & Sy1.9/LINER	 & 0.00507 	 &516   &  13 07 1996 00:03:03	\\
NGC4593	     &RSBb(rs)b      & Sy1		 & 0.00900	 &516   &  14 07 1996 06:20:04	\\
NGC5033	     &SA(s)c	     & Sy1.9		 & 0.00292	 &516   &  25 06 1996 12:13:25	\\
Mrk509	     & Compact	     & Sy1.2		 & 0.03440	 &516   &  18 10 1996 03:03:22	\\
NGC526A	     &S0pec?	     & Sy1.5		 & 0.01922	 &516   &  24 11 1996 03:13:59	\\
NGC5273	     &SA(s)0	     & Sy1.9		 & 0.00352	 &516   &  25 06 1996 11:51:23	\\
NGC5548	     &RSA(s)0/a      & Sy1.5		 & 0.01717 	 &514   &  07 02 1996 16:57:22	\\
NGC5674	     &SABc	     & Sy1.9		 & 0.02492	 &514   &  07 02 1996 14:05:36	\\
NGC5728	     &RSAB(r)a       & Sy2		 & 0.00930	 &514   &  07 02 1996 08:49:23	\\
Mrk590	     &SA(s)a	     & Sy1.2		 & 0.02638	 &516   &  03 07 1997 15:11:27	\\
NGC5929	     & Sab: pec      & Sy2		 & 0.00831	 &514   &  05 02 1996 13:31:51	\\
NGC5940	     &SBab	     & Sy1		 & 0.03405 	 &514   &  07 02 1996 15:06:16	\\
NGC5953	     &SAa pec	     & Sy2/LINER 	 & 0.00656	 &514   &  07 02 1996 16:24:58	\\
Mrk673	     & S?	     & Sy2/LINER/HII	 & 0.03651	 &514   &  07 02 1996 19:55:02	\\
NGC701	     &SB(rs)c	     & SB		 & 0.00610	 &516   &  18 12 1997 00:35:41	\\
NGC7314	     &SAB(rs)bc      & Sy1.9		 & 0.00474	 &516   &  29 04 1996 12:42:14	\\
NGC7592      & S0+pec        & Sy2		 & 0.02444	 &516   &  15 11 1996 15:56:53	\\
NGC7603	     &SA(rs)b pec    & Sy1.5	 	 & 0.02952	 &516   &  18 05 1996 05:47:11	\\
Mrk766	     &RSB(s)a	     & Sy1.5	 	 & 0.01293	 &516   &  02 06 1996 04:17:32	\\
NGC7674	     & SA(r)bc pec   & Sy2/HII  	 & 0.02906	 &516   &  28 05 1996 06:39:23	\\
Mrk789	     & Irregular     & Sy1/HII  	 & 0.03145	 &516   &  16 12 1996 12:51:04	\\
Mrk817	     & SBc	     & Sy1.5		 & 0.03145	 &514   &  05 02 1996 14:04:02	\\
Mrk841	     & E	     & Sy1.5		 & 0.03620 	 &514   &  07 02 1996 15:54:02	\\
A~1058+45    & Sa            & Sy2		 & 0.02908 	 &516   &  18 04 1996 09:51:17	\\
Ark~120	     & Sb/pec	     & Sy1		 & 0.03273	 &516   &  20 08 1997 13:05:38	\\
ESO137-G34   & SAB(s)0/a?    & Sy2		 & 0.00916	 &514   &  09 02 1996 08:31:18	\\
ESO141-G55   & Sc	     & Sy1		 & 0.03600	 &524   &  04 03 1996 18:04:15	\\
FAIRALL9     &S 	     & Sy1	 	 & 0.04702	 &516   &  06 05 1996 09:28:16	\\
H~1821+643   &  	     & QSO/Sy1   	 & 0.29700 	 &514   &  05 02 1996 10:11:07	\\
HS0624+6907  &  	     & QSO	 	 & 0.37000 	 &516   &  04 09 1997 09:06:18	\\
HS1700+6416  &  	     & QSO	 	 & 2.73574	 &826   &  18 10 1996 19:52:33	\\
IR~05189-2524& pec	     & Sy2		 & 0.04256	 &516   &  18 10 1997 00:57:35	\\
IR~12495-1308& Sa	     & Sy1		 & 0.01463	 &516   &  19 12 1996 11:29:20	\\
IR~22377+0747& SBa           & Sy1.8		 & 0.02460	 &516   &  18 05 1996 09:26:28	\\
MCG+8-11-11  & SB0	     & Sy1.5		 & 0.02048	 &516   &  14 10 1997 23:51:53	\\
MGC-6-30-15  &E-S0	     & Sy1.2		 & 0.00775	 &514   &  14 02 1996 17:51:30	\\
\hline
\end{tabular}
\caption{Morphological classification of the galaxies, Seyfert type and
spectroscopic redshift (obtained from the NASA/IPAC Extragalactic Database, $NED$).}
\label{info} 
\end{table}

\clearpage

\begin{deluxetable}{lcccccccccccc}
\tablecolumns{11}
\tablewidth{0pt}
\tabletypesize{\scriptsize}
\rotate
\tablecaption{Fitted parameters and errors for each galaxy in both LW2 and LW7 filters: PSF
intensity (in mJy), exponential component length scale, bar length scale, bar falling 
slope, ring radial offset, and ring width (all parameters except the PSF intensity
are given in pixels, being the pixel size of 3\arcsec). * Mrk~3, MGC-6-30-15 and 
Mrk~841 have been fitted by bulges (they appear in the bar parameters
columns in order to avoid enlargement of the table), bulge length scale and Sersic law index are given.
\label{params}} 
\tablehead{
\colhead{GALAXY} & \multicolumn{2}{c}{PSF int.} & \multicolumn{2}{c}{Exp. length scale} & \multicolumn{2}{c}{Bar length scale} & 
\multicolumn{2}{c}{Bar falling slope}&\multicolumn{2}{c}{Ring radial offset}&\multicolumn{2}{c}{Ring width} \\
\colhead{} & \colhead{LW2} & \colhead{LW7} & \colhead{LW2} & \colhead{LW7} & \colhead{LW2} & \colhead{LW7} & \colhead{LW2} & \colhead{LW7} &
\colhead{LW2} & \colhead{LW7} & \colhead{LW2} & \colhead{LW7}} 
\startdata
NGC1097	     &1.21$\pm$1.31&1.16$\pm$1.30&2.70$\pm$0.10&3.07$\pm$0.26 &\nodata&\nodata &\nodata &\nodata &3.41$\pm$0.11 & 3.27$\pm$0.08 & 0.75$\pm$0.11&0.79$\pm$0.18 \\
NGC1125	     &14.2$\pm$4.5&9.67$\pm$2.63&14.7$\pm$4.9&14.7$\pm$4.9&2.16$\pm$1.43 & 3.10$\pm$0.85&0.62$\pm$0.35&0.33$\pm$0.24&\nodata&\nodata&\nodata&\nodata    \\
NGC1144	     &\nodata&\nodata&7.59$\pm$2.11&7.50$\pm$2.10&2.98$\pm$0.56&2.88$\pm$0.08&0.06$\pm$0.02&0.06$\pm$0.05&\nodata & \nodata  &\nodata	   &\nodata \\
NGC1241	     &1.87$\pm$1.45&1.72$\pm$0.81&\nodata&\nodata&9.45$\pm$0.46&10.3$\pm$3.1&1.44$\pm$0.21&2.01$\pm$1.32&1.14$\pm$0.72&1.00$\pm$0.68&0.88$\pm$0.37&1.00$\pm$0.50 \\
Z~1335.5     &\nodata&\nodata&1.34$\pm$0.06 & 1.32$\pm$0.06 &	\nodata   &	\nodata    & \nodata	  &	\nodata   &\nodata & \nodata  &	\nodata   & \nodata\\
NGC1386	     &46.7$\pm$2.4&43.8$\pm$1.7&2.24$\pm$0.05&3.16$\pm$0.07&	\nodata   &	\nodata    &  \nodata	  &	\nodata   &\nodata & \nodata  &\nodata	   & \nodata\\
NGC1566	     &18.2$\pm$5.8&7.95$\pm$2.53&1.42$\pm$0.20 & 1.52$\pm$0.11&\nodata&\nodata&\nodata&\nodata&10.0$\pm$2.0&10.0$\pm$2.0 & 2.00$\pm$0.40 & 2.00$\pm$0.40  \\
NGC1667	     &\nodata&\nodata&2.10$\pm$0.20&2.62$\pm$0.62&7.63$\pm$0.64&8.71$\pm$0.65  &1.19$\pm$0.18 & 1.18$\pm$0.22&\nodata & \nodata  &	\nodata   &\nodata \\
Mrk266	     &13.3$\pm$5.4&13.7$\pm$5.3&1.21$\pm$0.13& 1.76$\pm$0.17 & \nodata    &	\nodata    &	\nodata  &	\nodata   &\nodata & \nodata  &	\nodata   & \nodata \\
Mrk279	     &22.2$\pm$2.0&29.3$\pm$6.7&1.84$\pm$0.10&1.18$\pm$0.09&	 \nodata  &	\nodata    &	\nodata  &\nodata	   &\nodata &  \nodata &\nodata	   &\nodata \\
NGC3227	     &58.2$\pm$6.7&54.6$\pm$5.5&1.29$\pm$0.15&1.30$\pm$0.20&0.40$\pm$0.10 & 0.40$\pm$0.10&13.2$\pm$2.6&13.0$\pm$3.0 &\nodata &  \nodata &\nodata	   & \nodata\\
Mrk334	     &26.8$\pm$3.0&17.1$\pm$3.9&0.99$\pm$0.06& 0.88$\pm$0.05&\nodata	   &	\nodata    &	\nodata  &	\nodata   &\nodata &  \nodata &\nodata	   &\nodata \\
Mrk335	     &43.3$\pm$4.8&34.6$\pm$5.0 &1.23$\pm$0.10 & 1.20$\pm$0.10 &	\nodata   &	\nodata    &	\nodata  &	\nodata   & \nodata& \nodata  &\nodata	   &\nodata \\
NGC3516	     &59.6$\pm$8.2&76.3$\pm$10.3 &1.45$\pm$0.11 & 1.42$\pm$0.09 &\nodata	   &	\nodata    &	\nodata  &	\nodata   &\nodata & \nodata  &	\nodata   &\nodata \\   
NGC3982	     &1.50$\pm$0.20&2.36$\pm$0.34&  \nodata   &	\nodata     &5.30$\pm$0.30&5.54$\pm$0.23&1.69$\pm$0.07& 1.75$\pm$0.05&\nodata & \nodata  &	\nodata   &\nodata \\
3C382	     &15.0$\pm$3.1&17.5$\pm$4.2&1.32$\pm$0.19& 2.02$\pm$0.41 &\nodata	   &	\nodata    &	\nodata  &\nodata	   &\nodata & \nodata  &	\nodata   & \nodata \\
Mrk3	     &\nodata&\nodata&\nodata&\nodata&(8.77$\pm$0.59)&(8.97$\pm$0.33)&(2.40$\pm$0.10)&(2.02$\pm$0.09) & \nodata& \nodata  &\nodata	   &\nodata \\
NGC4051	     &58.4$\pm$9.1&65.0$\pm$15.8&13.0$\pm$3.0& 12.9$\pm$3.4&0.30$\pm$1.10 &0.29$\pm$1.02&0.75$\pm$0.28&0.80$\pm$0.30&\nodata  & \nodata   &	\nodata    & \nodata \\
IC4329A	     &155$\pm$32&209$\pm$32&1.98$\pm$0.13 & 2.10$\pm$0.10 &	\nodata   &  \nodata      &  \nodata	  &   \nodata     & \nodata& \nodata  &	\nodata   & \nodata \\
NGC4388	     &44.2$\pm$3.6&38.8$\pm$4.9&19.6$\pm$4.5&20.0$\pm$5.0&2.50$\pm$0.40&2.20$\pm$0.30&0.30$\pm$0.10&0.30$\pm$0.10   &\nodata  &\nodata    &\nodata 	   &\nodata  \\
IC4397	     &2.37$\pm$0.26&   1.69$\pm$0.25&3.37$\pm$0.11& 3.70$\pm$0.10 &	\nodata    &	\nodata     &	\nodata   &	\nodata    &\nodata  & \nodata   &	\nodata    & \nodata  \\  
NGC4507	     &55.8$\pm$8.4&89.4$\pm$16.2&0.90$\pm$0.20&1.34$\pm$0.15&9.34$\pm$0.44&9.57$\pm$0.59&0.90$\pm$0.20&0.95$\pm$0.24 &\nodata & \nodata  &	\nodata   &\nodata \\  
NGC4579	     &12.0$\pm$1.3   &   16.4$\pm$1.7 &2.01$\pm$0.07 & 2.77$\pm$0.16 &	\nodata    &	\nodata     &	\nodata   &\nodata 	   &\nodata  & \nodata   &	\nodata    & \nodata  \\
NGC4593	     &32.9$\pm$13.1   &   43.5$\pm$8.0 &0.99$\pm$0.26 & 1.02$\pm$0.11 &	\nodata    &	\nodata     &	\nodata   &	\nodata    &\nodata  &  \nodata  &	\nodata    &\nodata  \\
NGC5033	     &4.05$\pm$0.49&4.38$\pm$0.44&\nodata &\nodata&5.54$\pm$0.24&5.92$\pm$0.33&0.87$\pm$0.15&1.04$\pm$0.18&9.13$\pm$0.34&9.60$\pm$0.30&3.06$\pm$0.16&2.70$\pm$0.10 \\		 
Mrk509	     &44.1$\pm$5.3&   42.8$\pm$7.8 &1.44$\pm$0.16 & 1.20$\pm$0.10 &	\nodata   &	 \nodata   &	\nodata  &	\nodata   & \nodata& \nodata  &	\nodata   & \nodata \\
NGC526A	     &32.8$\pm$2.7&35.2$\pm$5.9 &1.64$\pm$0.15 & 1.52$\pm$0.15 &\nodata 	   &	\nodata     &	\nodata   &	\nodata    &\nodata  & \nodata   &	\nodata    &\nodata  \\
NGC5273	     &2.96$\pm$0.46&2.90$\pm$0.80  &3.21$\pm$0.33 & 2.92$\pm$0.35 &	\nodata   &	\nodata    &\nodata	  &\nodata	   & \nodata& \nodata  &	\nodata   &\nodata \\
NGC5548	     &40.6$\pm$3.3&42.2$\pm$4.1 &1.58$\pm$0.08 & 1.49$\pm$0.07 &	\nodata    &	\nodata     &\nodata 	  &	\nodata    &\nodata  & \nodata   &	\nodata    & \nodata \\
NGC5674	     &1.97$\pm$1.47&2.59$\pm$1.93&1.97$\pm$0.19&1.28$\pm$0.31&\nodata  &\nodata  &\nodata  &\nodata &8.00$\pm$2.00 & 7.50$\pm$1.00 &1.00$\pm$0.10 & 2.00$\pm$0.10  \\
NGC5728	     &\nodata&\nodata&6.06$\pm$3.45 & 6.00$\pm$3.00 &0.60$\pm$0.50 & 0.60$\pm$0.50   &0.65$\pm$0.16 & 0.61$\pm$0.07   & \nodata& \nodata  &	\nodata   &\nodata \\
Mrk590	     &31.5$\pm$8.9   &   47.7$\pm$11.1 &2.67$\pm$0.30 & 3.52$\pm$0.40 &	 \nodata  &	\nodata    &	\nodata  &	\nodata   & \nodata& \nodata  &\nodata	   &\nodata  \\
NGC5940	     &2.38$\pm$0.26&1.58$\pm$1.04&32.0$\pm$1.0&32.5$\pm$0.7&4.97$\pm$0.21&3.19$\pm$2.20&0.06$\pm$0.17&1.50$\pm$0.70&\nodata  & \nodata   &	\nodata    &\nodata  \\
NGC5953	     &\nodata&\nodata&1.50$\pm$0.80 & 2.11$\pm$3.25 &2.43$\pm$0.86 & 2.18$\pm$1.12   &0.25$\pm$0.14 & 0.44$\pm$0.29   & \nodata&  \nodata &	\nodata   & \nodata\\
Mrk673	     &7.59$\pm$3.80&2.00$\pm$2.00&1.76$\pm$1.92&5.20$\pm$4.60&1.26$\pm$0.92 & 1.57$\pm$0.71&0.68$\pm$0.26 & 0.46$\pm$0.25   &\nodata  & \nodata   &\nodata 	   &\nodata  \\
NGC701	     &5.57$\pm$1.22 &   2.12$\pm$0.95  &4.18$\pm$0.16 & 4.55$\pm$0.19 &	\nodata    &	\nodata     &\nodata 	  &\nodata 	   &\nodata  & \nodata   &\nodata 	   & \nodata  \\
NGC7314	     &8.21$\pm$4.99&8.72$\pm$2.52  &7.30$\pm$4.30 & 7.00$\pm$4.00 &1.00$\pm$0.10 & 1.30$\pm$0.30&0.58$\pm$0.21 & 0.05$\pm$0.03   & \nodata& \nodata  &\nodata	   &\nodata \\
NGC7603	     &32.9$\pm$1.9&   29.3$\pm$1.7 &2.60$\pm$0.40 & 2.60$\pm$0.40 &	 \nodata  &	\nodata    &	\nodata  &\nodata	   &\nodata & \nodata  &\nodata	   &\nodata \\  
Mrk766	     &33.5$\pm$8.2&42.4$\pm$10.6 &4.50$\pm$1.90 & 4.54$\pm$1.89 &1.00$\pm$0.10 & 1.05$\pm$0.99   &0.60$\pm$0.20 & 0.61$\pm$0.24   &\nodata &  \nodata &	\nodata   & \nodata \\  
NGC7674	     &42.4$\pm$6.2&58.6$\pm$16.2 &2.13$\pm$0.15 & 2.51$\pm$0.28 &	 \nodata   &	\nodata     &	\nodata   &\nodata 	   &\nodata  &  \nodata  &	\nodata    &\nodata  \\
Mrk789	     &6.79$\pm$5.36&3.36$\pm$3.59  &6.00$\pm$8.00 & 5.89$\pm$7.68 &2.30$\pm$0.90 & 2.31$\pm$0.92&0.63$\pm$0.40 &0.63$\pm$0.40   & \nodata & \nodata   &	\nodata    & \nodata \\
Mrk817	     &27.7$\pm$3.9  &   53.1$\pm$4.9 &1.23$\pm$0.08 & 1.51$\pm$0.08 &	\nodata    &	 \nodata    &	\nodata   &	\nodata    &\nodata  & \nodata   &	\nodata    & \nodata  \\
Mrk841	     &\nodata&\nodata&\nodata&\nodata&(5.72$\pm$0.83)&(6.56$\pm$0.61)&(1.87$\pm$0.40)&(1.66$\pm$0.28)  & \nodata &  \nodata  &	 \nodata   & \nodata \\
A~1058+45    &1.96$\pm$0.37&   3.11$\pm$1.13  &2.94$\pm$0.17 & 3.18$\pm$0.23 &	 \nodata    &	 \nodata     & \nodata 	  &	 \nodata    &  \nodata &  \nodata   &	 \nodata    &  \nodata  \\
Ark~120	     &43.5$\pm$24.7&17.8$\pm$13.7 &1.59$\pm$0.60 & 1.41$\pm$0.18 &\nodata  	   &	\nodata      &\nodata  	  &\nodata  	   & \nodata  &  \nodata   & \nodata 	   & \nodata  \\	  
ESO141-G55   &22.6$\pm$2.9&21.7$\pm$11.4 &13.0$\pm$3.1& 15.0$\pm$4.0&0.61$\pm$1.13 & 0.50$\pm$0.90&0.85$\pm$0.30 &0.61$\pm$0.29   & \nodata  &  \nodata   & \nodata 	   &  \nodata \\
FAIRALL9     &47.3$\pm$9.2 & 53.5$\pm$9.9 &1.33$\pm$0.16 & 1.47$\pm$0.10 &	\nodata    &	\nodata     &	\nodata   &\nodata 	   & \nodata  &  \nodata   & \nodata 	   &  \nodata \\ 
H~1821+643   &21.2$\pm$1.6   &   27.4$\pm$2.1 &1.36$\pm$0.06 & 1.28$\pm$0.05 &	\nodata    &	\nodata     &	\nodata   &	\nodata    & \nodata  &  \nodata   & \nodata 	   & \nodata  \\  
HS0624+6907  &12.3$\pm$2.1   &   14.6$\pm$2.5 &1.35$\pm$0.14 & 1.47$\pm$0.15 &	\nodata    &	\nodata     &\nodata 	  &	\nodata    & \nodata  &  \nodata   & \nodata 	   & \nodata  \\  
IR~05189-2524&47.1$\pm$10.1&71.5$\pm$22.8&3.19$\pm$0.39 & 3.41$\pm$0.58 &1.85$\pm$0.13 & 1.87$\pm$0.85  &0.02$\pm$0.04 & 0.18$\pm$0.22   & \nodata  &  \nodata   & \nodata 	   & \nodata  \\
IR~12495-1308&7.25$\pm$0.56    &   10.7$\pm$1.1 &4.06$\pm$0.46 & 3.51$\pm$0.44 &	\nodata   &\nodata	    &\nodata	  &	\nodata   & \nodata  &  \nodata   & \nodata 	   & \nodata  \\  
IR~22377+0747&7.51$\pm$3.93&6.30$\pm$3.30&0.89$\pm$0.32&0.65$\pm$1.82&8.95$\pm$3.11 & 9.11$\pm$2.17&0.29$\pm$1.65 & 0.56$\pm$1.21 & \nodata  &  \nodata   & \nodata 	   & \nodata  \\
MCG+8-11-11  &54.7$\pm$14.7&47.9$\pm$21.0 &0.84$\pm$0.18 & 0.73$\pm$0.11 &12.3$\pm$0.5& 12.5$\pm$0.5 &0.54$\pm$0.73 &0.53$\pm$0.72   & \nodata  &  \nodata   & \nodata 	   & \nodata  \\  
MGC-6-30-15  &43.1$\pm$16.3&54.1$\pm$11.5&\nodata&\nodata&(7.35$\pm$0.75)&(8.13$\pm$0.37)&(1.12$\pm$0.27)&(1.00$\pm$0.10) & \nodata  & \nodata    & \nodata 	   & \nodata  \\	  
\enddata
\end{deluxetable}

\clearpage

\begin{deluxetable}{l c c c c c c c c c c}
\tablecolumns{11}
\tablewidth{0pt}
\tabletypesize{\scriptsize}
\rotate
\tablecaption{Fluxes (in mJy) obtained by an integration of the emission of each individual component
(PSF profile, exponential component, bar and ring) plus the total fit. For Mrk~3, MGC6-30-15, and Mrk~841, 
which have been fit to bulges, total fluxes are given, like in the case of ESO137-G34, NGC5929, NGC7592, and HS1700+6416, whose 
total fluxes were obtained by aperture photometry. 
\label{fluxes}} 
\tablehead{
\colhead{GALAXY} & \multicolumn{2}{c}{PSF flux} & \multicolumn{2}{c}{Exp. comp. flux} & \multicolumn{2}{c}{Bar flux} & 
\multicolumn{2}{c}{Ring flux} & \multicolumn{2}{c}{Total flux} \\
\colhead{} & \colhead{LW2} & \colhead{LW7} & \colhead{LW2} & \colhead{LW7} & \colhead{LW2} & \colhead{LW7} & \colhead{LW2} & \colhead{LW7} &
\colhead{LW2} & \colhead{LW7}} 
\startdata
NGC1097	     &1.70$\pm$1.60  &3.10$\pm$2.90&448$\pm$10&490$\pm$23& \nodata & \nodata&543$\pm$113&555$\pm$86&1008$\pm$29&1054$\pm$32 \\
NGC1125	     &38.1 $\pm$6.3&25.5 $\pm$4.4&7.82$\pm$0.45&28.0$\pm$1.0&26.0$\pm$21.0&31.0$\pm$25.0& \nodata & \nodata &70.4 $\pm$9.6&83.0$\pm$5.0 \\
NGC1144	     & \nodata & \nodata &89.0$\pm$5.0&86.2$\pm$1.4&70.0$\pm$4.0&62.0$\pm$9.0& \nodata & \nodata & 166$\pm$4 & 155$\pm$5 \\
NGC1241	     &4.14$\pm$2.27&2.40$\pm$0.90&\nodata &\nodata&155$\pm$21&166$\pm$80&22.0$\pm$9.0&17.4$\pm$1.9&179$\pm$11&197$\pm$10 \\
Z~1335.5+3925&\nodata &\nodata &47.4$\pm$2.8&56.7$\pm$2.8& \nodata &\nodata & \nodata& \nodata&49.0$\pm$3.1&57.3$\pm$2.8 \\
NGC1386	     &125$\pm$5&117$\pm$3&120$\pm$2&172$\pm$2&  \nodata  &  \nodata &  \nodata &\nodata &246$\pm$8&287$\pm$5 \\
NGC1566	     &48.6$\pm$11.3&11.6$\pm$2.7&73.0$\pm$11.0&128$\pm$9&\nodata&\nodata&29.1$\pm$1.3&113$\pm$8&147$\pm$13&265$\pm$19 \\
NGC1667	     &\nodata & \nodata&47.0$\pm$8.0&69.0$\pm$10.0&260$\pm$24&262$\pm$21&\nodata & \nodata&307$\pm$10&334$\pm$11 \\
Mrk266	     &36.0$\pm$10.0&36.0$\pm$9.0&78.0$\pm$16.0&89.0$\pm$18.0&\nodata&\nodata&\nodata&\nodata&114$\pm$14&126$\pm$13 \\
Mrk279	     &59.4$\pm$3.3&79.0$\pm$11.0&22.7$\pm$1.4&54.0$\pm$7.0&\nodata&\nodata&\nodata & \nodata&81.3$\pm$4.6&133$\pm$14 \\
NGC3227	     &156$\pm$13&146$\pm$12&80.0$\pm$14.0&140$\pm$5&48.0$\pm$7.0&57.6$\pm$0.7&\nodata&\nodata &278$\pm$14&355$\pm$17 \\
Mrk334	     &72.0$\pm$5.0&46.0$\pm$8.0&40.0$\pm$5.0&77.0$\pm$10.0&\nodata&\nodata &\nodata & \nodata&111$\pm$7&128$\pm$9 \\
Mrk335	     &117$\pm$9&93.0$\pm$10.0&31.0$\pm$4.0&81.0$\pm$8.0&\nodata&\nodata& \nodata&\nodata &144$\pm$12&176$\pm$13 \\
NGC3516	     &160$\pm$15&203$\pm$20&66.0$\pm$5.0&121$\pm$8&\nodata&\nodata&\nodata & \nodata&228$\pm$20&332$\pm$26 \\
NGC3982	     &2.14$\pm$0.26&4.31$\pm$0.53&\nodata&\nodata&279$\pm$11&306$\pm$9& \nodata&\nodata &297$\pm$10&327$\pm$7 \\
3C382	     &40.0$\pm$5.7&46.0$\pm$12.0&18.0$\pm$3.0&21.0$\pm$15.0&\nodata&\nodata& \nodata& \nodata&57.2 $\pm$7.3&67.3 $\pm$10.6 \\
Mrk3	     &\nodata &    \nodata   &   \nodata   &  \nodata&      \nodata&   \nodata&  \nodata &  \nodata&134$\pm$7&245$\pm$15 \\
NGC4051	     &156$\pm$17&174$\pm$34&45.7$\pm$0.8&51.0$\pm$8.0&\nodata  &    \nodata&  \nodata &  \nodata &280$\pm$23&361$\pm$30 \\
IC4329A	     &418$\pm$45&563$\pm$45&116$\pm$14&203$\pm$15&\nodata &\nodata&  \nodata &  \nodata &513$\pm$65&753$\pm$66 \\
NGC4388	     &117$\pm$8&103$\pm$18&192$\pm$9&183$\pm$3&36.0$\pm$12.0&65.0$\pm$21.0&  \nodata & \nodata  &342$\pm$11&350$\pm$16  \\
IC4397	     &6.33$\pm$0.42&4.50$\pm$0.60&60.4$\pm$1.3&93.7 $\pm$2.2&\nodata&\nodata &\nodata & \nodata&67.1$\pm$1.1&99.9$\pm$2.1 \\
NGC4507	     &148$\pm$17&241$\pm$35&92.0$\pm$16.0&111$\pm$20&71.9$\pm$3.2&68.0 $\pm$6.0& \nodata  & \nodata  &318$\pm$17&404$\pm$25 \\
NGC4579	     &31.8$\pm$2.2&43.4 $\pm$3.7&71.1$\pm$2.3&91.0 $\pm$3.0 &\nodata &\nodata&\nodata&\nodata &103$\pm$3&135$\pm$5 \\
NGC4593	     &87.0$\pm$23.0&117$\pm$21&108$\pm$29&128$\pm$28&\nodata&\nodata&  \nodata & \nodata  &194$\pm$17&240$\pm$14 \\
NGC5033	     &10.5$\pm$0.6&11.5 $\pm$0.5& \nodata  & \nodata&220$\pm$6&205$\pm$7&364$\pm$9&325$\pm$11&594$\pm$5&541$\pm$4\\
Mrk509	     &118$\pm$8&115$\pm$11&34.0$\pm$4.0& 83.0$\pm$10.0 &    \nodata &\nodata&  \nodata &\nodata &151$\pm$12&194$\pm$16 \\
NGC526A	     &88.4 $\pm$4.5&94.0 $\pm$12.0&23.0 $\pm$2.0&51.0 $\pm$4.0&    \nodata&\nodata& \nodata&\nodata&108$\pm$7&144$\pm$11 \\
NGC5273	     &7.91  $\pm$0.95&7.60  $\pm$1.50&14.9 $\pm$0.9&25.0 $\pm$2.0&\nodata&\nodata& \nodata&\nodata &23.5 $\pm$1.3&34.0 $\pm$2.3 \\
NGC5548	     &109$\pm$6&113$\pm$7&43.8 $\pm$2.6&102$\pm$6&   \nodata&     \nodata& \nodata&\nodata &150$\pm$8&225$\pm$11 \\
NGC5674	     &2.92$\pm$1.41&6.88  $\pm$3.32&51.0$\pm$5.0&40.0$\pm$8.0&\nodata &\nodata &38.0$\pm$6.0&63.3$\pm$2.2&92.0$\pm$9.0&113$\pm$5 \\
NGC5728	     &\nodata&\nodata&41.0$\pm$6.0&122$\pm$2&121$\pm$48&74.1$\pm$5.9& \nodata&\nodata &162$\pm$12&200$\pm$10 \\
Mrk590	     &84.0 $\pm$18.0&124$\pm$21&57.0 $\pm$7.0&181$\pm$14&\nodata& \nodata  &\nodata &\nodata &146$\pm$23&303$\pm$25 \\
NGC5929	     &\nodata	& \nodata & \nodata & \nodata &\nodata  &  \nodata & \nodata & \nodata &12.4$\pm$0.1&13.0$\pm$0.1			\\
NGC5940	     &5.24  $\pm$0.77&4.17$\pm$1.68&8.46  $\pm$0.66&28.7 $\pm$1.1&25.0$\pm$5.0&46.0$\pm$14.0&\nodata&\nodata&39.4$\pm$1.9&81.0$\pm$5.0\\
NGC5953	     &\nodata& \nodata  &86.0$\pm$28.0&80.0$\pm$48.0&186$\pm$79&143$\pm$70&\nodata& \nodata&272$\pm$4&223$\pm$2 \\
Mrk673	     &20.0$\pm$9.0&4.10$\pm$1.90&2.72$\pm$2.04&21.0$\pm$7.0&30.0$\pm$11.0&49.0$\pm$21.0&\nodata&\nodata&52.3$\pm$6.0&75.0 $\pm$7.0 \\
NGC701	     &8.24$\pm$1.28&3.09$\pm$0.98&210$\pm$4&212$\pm$5&\nodata&\nodata&\nodata&\nodata&217$\pm$5&215$\pm$5 \\
NGC7314	     &18.0$\pm$6.0&23.0$\pm$6.0&112$\pm$19&104$\pm$3&33.0$\pm$8.0&22.0$\pm$5.0&\nodata &\nodata &166$\pm$10&163$\pm$11 \\
NGC7592	     &	\nodata& \nodata & \nodata & \nodata & \nodata &  \nodata & \nodata & \nodata &161$\pm$1&143$\pm$1		\\	
NGC7603	     &87.9$\pm$2.2&78.5$\pm$10.1&28.8$\pm$0.4&90.0$\pm$11.0&\nodata&\nodata&\nodata &\nodata&114$\pm$4&168$\pm$4 \\
Mrk766	     &89.6$\pm$10.8&113$\pm$18&21.3$\pm$0.6&54.0$\pm$17.0&55.9$\pm$5.2&95.0$\pm$9.0&\nodata &\nodata &161$\pm$17&260$\pm$23	 \\
NGC7674	     &113$\pm$10&157$\pm$30&147$\pm$8&222$\pm$18&\nodata&\nodata&\nodata &\nodata &267$\pm$19&376$\pm$41 \\
Mrk789	     &15.3 $\pm$7.3&6.00$\pm$4.00&7.76 $\pm$0.73& 22.0$\pm$9.0&51.0$\pm$6.0&47.0$\pm$14.0& \nodata& \nodata&76.0$\pm$11.0&75.0 $\pm$8.0\\
Mrk817	     &75.0$\pm$7.0&142$\pm$9&45.0$\pm$3.0&81.0 $\pm$3.0 &  \nodata            &    \nodata & \nodata& \nodata&123$\pm$10&226$\pm$12 \\
Mrk841	     &	 \nodata & \nodata&\nodata &  	\nodata     &   \nodata & \nodata&\nodata &\nodata &67.6 $\pm$7.0&119$\pm$7 \\           
A~1058+45    &5.17$\pm$0.79&8.19 $\pm$1.78&31.6$\pm$1.7&45.3 $\pm$2.1&\nodata&\nodata& \nodata& \nodata&37.0$\pm$1.7&54.6$\pm$3.7 \\
Ark~120	     &96.0$\pm$33.0&26.0$\pm$13.0&73.0$\pm$29.0&147$\pm$26&  \nodata& \nodata& \nodata& \nodata&171$\pm$37&180$\pm$27\\
ESO137-G34   &\nodata	&\nodata  & \nodata & \nodata & \nodata & \nodata  & \nodata & \nodata &138$\pm$1&94.6$\pm$0.5		\\
ESO141-G55   &61.0 $\pm$6.0&59.0$\pm$18.0 &4.75 $\pm$1.47&6.00$\pm$1.00 &30.0$\pm$12.0&83.0$\pm$35.0& \nodata&\nodata &93.7$\pm$5.9&136$\pm$20 \\
FAIRALL9     &126$\pm$15&144$\pm$16&40.0$\pm$7.0&85.0$\pm$12.0&\nodata&\nodata& \nodata&\nodata &169$\pm$24&229$\pm$23 \\
H~1821+643   &57.1 $\pm$2.9&73.4 $\pm$9.8&32.6 $\pm$2.4&54.0 $\pm$7.0&   \nodata&\nodata& \nodata&\nodata &91.3$\pm$3.6&125$\pm$5 \\
HS0624+6907  &33.1 $\pm$3.8&39.0 $\pm$4.0&12.9 $\pm$0.7&31.0 $\pm$3.0& \nodata & \nodata   & \nodata& \nodata&46.6 $\pm$5.2&63.3 $\pm$3.2 \\
HS1700+6416  & \nodata& \nodata& \nodata&\nodata &\nodata & \nodata&\nodata &\nodata &3.30$\pm$0.10&4.80$\pm$0.10 \\
IR~05189-2524&127$\pm$14&193$\pm$42&62.0$\pm$5.0&78.0$\pm$4.0&68.0$\pm$13.0&134$\pm$26&\nodata &\nodata &253$\pm$18&411$\pm$42\\
IR~12495-1308&19.2 $\pm$0.9&28.6$\pm$1.9&30.9$\pm$1.6& 53.8$\pm$3.5&  \nodata&\nodata& \nodata&\nodata &49.4 $\pm$1.4&84.0$\pm$3.0 \\
IR~22377+0747&20.1 $\pm$5.8&16.8$\pm$4.9&18.0$\pm$3.0&28.0$\pm$4.0 &8.02 $\pm$1.97&22.1$\pm$3.3&\nodata &\nodata&43.7$\pm$8.7&66.0 $\pm$19.0 \\
MCG+8-11-11  &147$\pm$29&127$\pm$44&61.0$\pm$9.0&179$\pm$18&38.3$\pm$1.8&65.9$\pm$1.2& \nodata& \nodata&242$\pm$21&368$\pm$33 \\
MGC-6-30-15  &115$\pm$36&144$\pm$24&  \nodata &  \nodata &    \nodata &   \nodata  &\nodata &\nodata &196$\pm$13&272$\pm$ 11\\              
\enddata
\end{deluxetable}

\end{document}

\title{Erratum: "The mid--infrared emission of Seyfert galaxies.
 A new analysis of ISOCAM data\footnote{Based on observations with ISO, an 
ESA project with instruments funded by ESA Member States 
(especially the PI countries: France, Germany, the Netherlands and 
the United Kingdom) and with the participation of ISAS and NASA.} "(AJ, 134,
2006[2007])}

\author{C. Ramos Almeida\altaffilmark{1}, A.M. P\'{e}rez Garc\'\i a\altaffilmark{1},
J.A. Acosta-Pulido\altaffilmark{1}, and J.M. Rodr\'\i guez Espinosa\altaffilmark{1}}

\altaffiltext{1}{Instituto de Astrof\'\i sica de Canarias (IAC), 
              C/V\'\i a L\'{a}ctea, s/n, E-38200, La Laguna, Tenerife, Spain.
cra@iac.es, apg@iac.es, jap@iac.es, and jre@iac.es}

	In our original paper (Section 4.4) we compared the ratios of hard X--ray to nuclear 
mid--infrared emission for Seyfert 1 and 2 nuclei, finding that both distributions appear 
significatively different according to the Kolmogorov-Smirnov test. 
We have detected a mistake in the computation of the X-ray luminosities 
(the lack of a 4$\pi$ factor). 
Additionally, for two galaxies, the nuclear mid-infrared fluxes used within calculations are slightly 
different from the correct values which are reported in Table 3 of the original version. 
These mistakes affect the mentioned ratios of hard X--ray to nuclear mid-infrared emission, although
does not alter the result concerning the distributions being statistically different.
The correct values of these ratios are  
$<log (L^X_{intr}/L^{MIR})> = -0.21\pm0.33$ for the Seyfert 1 galaxies, 
and $<log (L^X_{intr}/L^{MIR})> = 0.17\pm0.62$ for the type 2 nuclei. 
Figure \ref{lutz_nuc_lumin} has been updated with the newly computed values.
%for the hard X-ray luminosity
We have also corrected the X-axis respect to
the original figure, now being $\nu$L$_{\nu}$. We have performed again the Spearman's rank 
correlation test, finding that the
correlation is significant when all galaxies and types 1 and 2 are considered separately.
It is worth to note that here we have excluded from the analysis two 
Seyfert 2 galaxies (NGC~1386 and NGC~7674) due to their extremely low hard X-ray luminosities, 
consistent with being optically thick sources. 
Recently, \citet{Horst07} have estimated an intrinsic X-ray luminosity for NGC~7674 of 
$log~L_{intr}^{X} = 44.56\  \mathrm{erg}\  \mathrm{s}^{-1}$, in contrast
to the non absorption-corrected value reported by \citet{Lutz04} 
($log~L_{obs}^{X} = 41.91\ \mathrm{erg}\  \mathrm{s}^{-1}$).  
Such a difference of about 2.5~dex corroborates our decision of do not include these galaxies. 

	In a recent job, \citet{Horst07} have argued that our luminosity ratios are $\sim$8 
larger than what they have found for their well-resolved objects, claiming
that our nuclear data are heavily contaminated by nuclear star formation. 
Instead, our new values are now of the same order of 
magnitude than theirs, even smaller for the case of Seyfert 2 galaxies. 
Therefore, we believe that despite the limited resolution of ISOCAM images, 
our nuclear fluxes are representative of the torus emission, and the 
contamination by circumnuclear star formation, although existing, is not dominant. 

\addtocounter{figure}{7}

\begin{figure}[!h]
\centering
\includegraphics[width=8cm,angle=90]{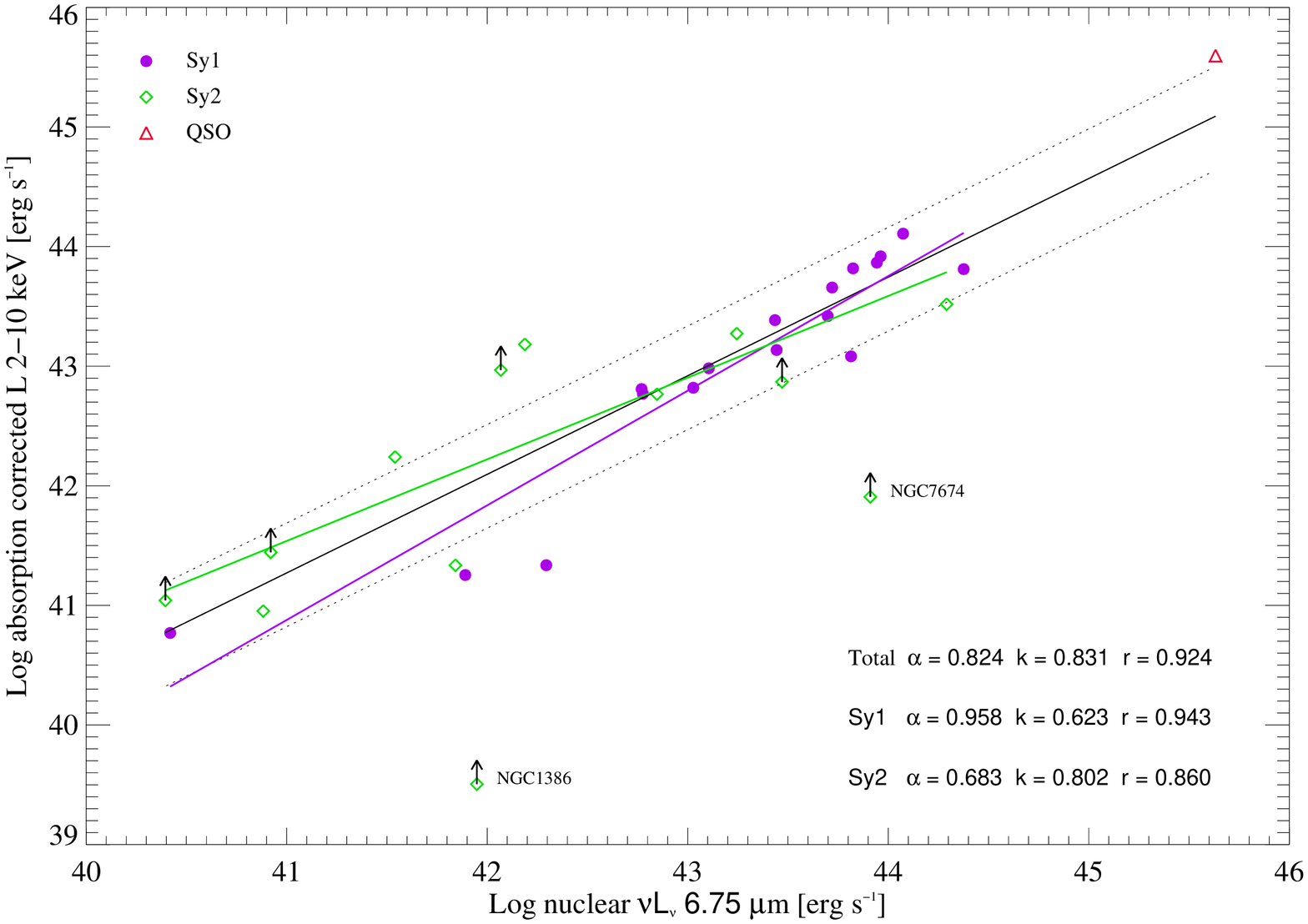}
\figcaption{\footnotesize{Absorption corrected hard X-ray luminosities versus nuclear
6.75~\micron\ luminosities. 
Solid line in black is the total fit and dotted lines correspond to one $\sigma$ limits. 
Filled circles represent Sy1 (17 in total), open diamonds Sy2 
(13 in total), and the QSO H~1821+643 is represented by an open triangle. 
Fits to Sy1 and Sy2 data are represented too.
Observed hard X--ray luminosities are represented as lower limits when no N$_{H}$ data is reported.
The galaxies NGC~7674 and NGC~1386 have not been taken into account in the correlation analysis.
The values of $\alpha$ and k are derived from the expression L$_{intr}^{X}/10^{43} = k(\nu L^{MIR}_{\nu}/10^{43})^{\alpha}$.} 
\label{lutz_nuc_lumin}}
\end{figure}

	There is a typographical mistake in Table 3 for the Point Spread Function (PSF) flux in the LW7 
filter corresponding to the galaxy ESO144-G55, the correct value being 84.2$\pm$2.44~mJy.